\documentclass{aa} 
\usepackage{graphicx}
\usepackage{amsmath}
\usepackage{longtable,lscape}

\usepackage{txfonts}
\begin{document}

  \title{FERO: Finding extreme relativistic objects}
  \subtitle{I. Statistics of relativistic Fe K$_\alpha$ lines in radio-quiet Type 1 AGN}

   \author{}
   \author{I. de la Calle P\'erez,
          \inst{1}
	  A.L. Longinotti,
	  \inst{2,1}
	  M. Guainazzi,
	  \inst{1}
          S. Bianchi,
	  \inst{3}
	  M. Dov\v{c}iak,
	  \inst{4}
          M. Cappi,
	  \inst{11}
	  G. Matt,
          \inst{3}
	  G. Miniutti,
	  \inst{5}
	  P.O. Petrucci,
	  \inst{6}
	  E. Piconcelli,
	  \inst{7}
          G. Ponti
	  \inst{8,9},
	  D. Porquet
	  \inst{10}
	  \and
	  M. Santos-Lle\'o
          \inst{1}
        }

   \offprints{I. de la Calle P\'erez}

   \institute{European Space Astronomy Centre of ESA, Apartado 50727, E-28080
         Madrid, Spain
	 \email{icalle@sciops.esa.int}
         \and
	  MIT Kavli Institute for Astrophysics and Space Research, Cambridge, USA
         \and
         Dipartimento di Fisica, Universit\`a degli Studi Roma Tre, via della
         Vasca Navale 84, 00146 Roma, Italy
	 \and
         Astronomical Institute AS CR, Bo\v{c}n\'{i} II 1401/1a, CZ-14131
         Praha 4, Czech Republic 
         \and
	 Centro de Astrobiolog\'ia (CSIC-INTA); LAEFF, P.O. Box 78, Villanueva de la Ca\~nada, Madrid, E-28691, Spain
         \and
	 Laboratoire d'Astrophysique, UMR5571 Universit\'e J. Fourier/CNRS, Observatoire de Grenoble BP53, F-38041 Grenoble cedex 9, France 
	 \and
	 Osservatorio Astronomico di Roma, Via Frascati 33, I-00040
         Monteporzio Catone, Italy 
         \and 
	 School of Physics and Astronomy, University of Southampton, Highfield,
	 Southampton SO17 1BJ, UK
         \and
          APC Universit\'e Paris 7 Denis Diderot, 75205 Paris, France
	 \and
	 Observatoire Astronomique de Strasbourg, Universit\'e Louis-Pasteur,
         CNRS, INSU, 11 rue de l'Universit\'e, 67000 Strasbourg, France
	 \and
	 INAF-IASF Bologna, via Gobetti 101, 40129 Bologna, Italy 
   }

   \date{Received December, 2009; accepted ?, 2009}


  \abstract {Accretion models predict that fluorescence lines broadened by
  relativistic effects should arise from reflection of X-ray emission onto the
  inner region of the accretion disc surrounding the central black hole of
  active galactic nuclei (AGN). The theory behind the origin of relativistic
  lines is well established, and observational evidence from a moderate number
  of sources seems to support the existence of these lines.}  
  {The aim of this work is to
  establish the fraction of AGN with relativistic Fe K$_\alpha$ lines, and study
  possible correlations with source physical properties.}  
  {An XMM-Newton collection of 149 radio-quiet Type 1 AGN has been
  systematically and uniformly analysed in order to search for 
  evidence of a relativistically broadened Fe K$_\alpha$ line. To enable statistical studies,
  an almost complete, flux-limited subsample of 31 sources has been defined by
  selecting the FERO sources observed by the RXTE all-sky Slew Survey with a
  count rate in the 3-8~keV energy band greater than 1 cts/sec. The 2-10~keV
  spectra of the FERO sources where compared with a complex model
  including most of the physical components observed in the X-ray spectra of
  Seyfert galaxies: a power law primary continuum modified by non-relativistic
  Compton reflection and warm absorption, plus a series of narrow Fe line
  reflection features.}
  {The observed fraction of sources in the
  flux-limited sample that show strong evidence of a relativistic Fe K$_\alpha$
  line is 36\%. This number can be interpreted as a lower limit to the fraction of
  sources that present a relativistic broad Fe K$_\alpha$ line in the wider AGN population. The average
  line equivalent width (EW) is of the order of 100~eV. The outcome of the fit
  yields an average disc inclination angle of 28 $\pm$ 5$^\circ$ and an average
  power-law index of the radial disc emissivity law of 2.4 $\pm$ 0.4. The spin
  value is well constrained only in 2 cases (MCG-6-30-15 and MRK~509); in
  the rest of the cases, whenever a constraint can be placed, it always
  implies the rejection of the static black hole solution. The Fe K$_\alpha$
  line EW does not correlate with disc parameters or with system physical
  properties, such as black hole mass, accretion rate, and hard X-ray luminosity.}  
  {} 

   \keywords{quasars: emission lines - galaxies: nuclei - galaxies: active -
               X-ray: galaxies - line: profiles
               }
   \authorrunning{I. de la Calle P\'erez et al.}
   \titlerunning{Statistics of Relativistic Fe K$_\alpha$ lines in Type 1 AGN}
   \maketitle
%

\section{Introduction}\label{Introduction}

\par The standard scenario for the X-ray emission in active galactic nuclei
(AGN) assumes that the observed X-ray power-law continuum originates in the
inner regions of the AGN closest to the central super-massive back hole, via
inverse Compton scattering of soft-energy photons in a corona of relativistic
electrons located somewhere above the accretion disc
(e.g. \cite{Haardt1993}). The X-ray illumination of optically thick cold matter,
such as the molecular torus and/or the accretion disc, gives rise to a Compton
reflection spectral component containing a series of fluorescent lines from
K$_\alpha$ transitions in metal atoms (e.g. \cite{George1991}). From a
combination of fluorescence yield (proportional to the fourth power of the
atomic number) and cosmic abundance, the most prominent is the K$_\alpha$
line emitted by neutral iron at 6.4~keV.

 \par When the Compton reflection originates in distant material like the
molecular torus envisaged in unification models (\cite{Antonucci1993}), the
profile of the Fe K$_\alpha$ emission line is narrow and unresolved by present
X-ray detectors. In contrast, the reflection component originating in the
inner accretion disc is affected by the black hole's strong gravitational
field, which modifies the line profile. The numerous studies of the effect of
gravity on the narrow emission line all agree by describing the resulting {\it
relativistic profile} as skewed and asymmetric because of a combination of
kinematic and relativistic effects (e.g. \cite{Fabian2000}), with a red wing
extending towards low X-ray energies due to gravitational redshift and
transverse Doppler redshift (e.g. \cite{Fabian1989}).

\par Since relativistic lines originate within a few gravitational radii from
the central object, the study of their shape and intensity may represent a
potential probe of the physical processes taking place in the innermost regions of
the AGN. The line profile is in fact very sensitive to the accretion disc
properties, such as the radial extension and dependence of the line
emissivity, the ionisation state of the material, the observer's inclination
angle relative to the disc axis, and the spin of the black hole (for a review
see \cite{Reynolds2003}, \cite{Fabian2005} and \cite{Matt2006} and references therein).

\par The first unambiguous observational evidence of relativistic Fe
K$_\alpha$ lines in the X-ray spectra of AGN was found in the spectrum of
MCG-6-30-15 obtained by the {\it ASCA} satellite (\cite{Tanaka1995}) and
lately, in several other Seyfert Galaxies (\cite{Nandra1997}) to the point
where the broad line was considered a common feature of AGN. Nevertheless, the presence
of relativistic broad Fe K$_\alpha$ lines is nowadays more controversial than
ever since only a handful of sources seem to possess a truly relativistically
broadened Fe~K$_\alpha$ line, e.g. MCG-6-30-15 (\cite{Fabian2002}) or
NGC~3516 (\cite{Markowitz2008}). When studies based on sizable samples of
sources are considered, the average fraction of sources with relativistic
broad Fe K$_\alpha$ lines is never higher than 40\% (\cite{Porquet2004},
\cite{Bailon2005}, \cite{Guainazzi2006}, \cite{Nandra2007}).

\par In the past decade the advent of {\it Chandra}, {\it XMM-Newton} and {\it
Suzaku} has undoubtedly provided deeper knowledge and more thorough
understanding of AGN X-ray spectral properties by means of superior spectral
resolution and higher throughput. For instance, it has been confirmed that
about half of quasars and Seyfert galaxies show the so-called warm absorbers,
i.e., ionised material outflowing along the line of sight, which is revealed
as a series of absorption features imprinted mainly in the soft X-ray spectral
band (\cite{Piconcelli2005}; \cite{Blustin2005} and references
therein). Emission lines from highly ionised atoms, including iron, can also
be observed (see \cite{Bianchi2005}). The presence of this absorbing/emitting
gas may introduce further complexities into the Fe K band: it has been shown
that warm absorbers with sufficiently high column density and ionisation state
may distort the continuum underlying the iron line, mimicking a relativistic
broad red wing (e.g. NGC~3783, \cite{Reeves2004}, and MCG-6-30-15,
\cite{Miller2008}; see also \cite{Turner2009} for a review). Also, a blend of
emission lines from helium and hydrogen-like iron at 6.7-6.97~keV can be
mistaken for a broad disk line observed at high inclination (NGC~7314,
\cite{Yaqoob2003}, and MRK~590, \cite{Longinotti2007}). While the latter
scenario is fully taken into account in our baseline model, only a simple
parameterization of the warm absorber was adopted, because an unknown number
of {\it a priori} absorbing systems do not allow a systematic analysis to be
performed (see Section~\ref{SpectralModel} for details).

\par The FERO (finding extreme relativistic objects) project is part of a
wider investigation on AGN carried out on archival {\it XMM-Newton} data (see
Section~\ref{Sample}). It was designed to address two fundamental questions on
the relativistic broad Fe K$_\alpha$ line:
\begin{itemize}
\item[i)] how common is relativistic broadening in AGN, and
\item[ii)] does the presence of a broad
line depend on other source's physical properties ?
\end{itemize}
To try to answer these questions, all the sources in our sample were fitted
with one and the same model, which includes all absorption and emission
components known to be potentially present in AGN (see
Section~\ref{SpectralModel}) and able to affect the emission in the iron K
band.

\par This paper reports the results of the spectral analysis on the individual
sources of the sample. A companion paper (Longinotti et al. in preparation) is
devoted to the analysis of stacked spectra. The structure of this paper is as
follows. Section~\ref{Sample} describes the selection of the
sample. Section~\ref{Analysis} describes the analysis procedure, including the
spectral analysis and the spectral model used to describe our
data. Section~\ref{Results} summarizes our results to be discussed in
Section~\ref{Discussion}. Section~\ref{Conclusions} closes the paper with the
conclusions drawn from the study presented here.


\section{The Sample}\label{Sample}

\par The FERO AGN collection proceeds from the CAIXA catalogue of AGN recently
published by Bianchi et al. (2009a,b). CAIXA is the largest catalogue of high
signal-to-noise X-ray spectra of AGN which consists of all the radio-quiet
X-ray unobscured ($N_{\rm H}<2\times10^{22}$cm$^{-2}$) AGN observed by {\it
XMM-Newton} in targeted observations, whose data are public as of March
2007. The sample, through a complete and homogeneous spectral analysis, is
characterized in terms of the parameters adopted by the best-fit models. CAIXA
includes a total of 77 quasars and 79 Seyfert galaxies. The redshift
distribution spans from z=0.002 to z=4.520 (almost 90\% within z$<$1) and the
distribution of the hard X-ray luminosities covers a range between
L$_{2-10keV}$ = 2.0 10$^{41}$ - 3.9 10$^{46}$ erg s$^{-1}$. Since the source
selection criteria used in this work are by large the same described there,
the reader is deferred to these works for more details. Here, those general
aspects that are more relevant to the relativistic line analysis carried out
in this work are recalled.

\subsection{The FERO AGN Collection}\label{FEROSample}

\par The starting sample consisted of 161 radio-quiet X-ray unobscured sources
targeted by {\it XMM-Newton} with public data up to April 2008. Only sources
with local column density from cold gas lower than N$_{\rm H}$ $\leq$ 2
$\times$ 10$^{22}$ cm$^{-2}$ are included in the sample to avoid heavily
absorbed spectra in the 2 to 10~keV spectral region. With respect to CAIXA
(which sums up 156 sources), the following 5 AGNs were added: ESO~511-G030,
IRAS~05078+1626, MRK~704, NGC~3227, NGC~526A. The first three sources were
specifically proposed and granted in the {\it XMM-Newton} AO6 for the purpose
of the FERO project; the other two objects became publicly available in the
data archive during the development of the FERO analysis. NGC~4151 fullfills
the required conditions to be in our sample, but it was excluded (both from
FERO and CAIXA) based on its complex X-ray spectra and extreme spectral
variability attributed to several absorbing systems
(\cite{Puccetti2007}). Based on a signal-to-noise criterion, as reasoned in
Section~\ref{SpectralAnalysis}, 12 sources are dropped from the starting
sample, leaving what constitutes the FERO sample with a total of 149 sources.

\par Out of the 149 sources, 67 are classified as quasars (RQQs) and 82
classified as Seyfert 1s (Sy) according to the value of the absolute optical
magnitude M$_B$ (QSO M$_{\rm B}$ $<$ -23; Sy M$_{\rm B}$ $>$ -23) as defined in the
V\'eron-Cetty $\&$ V\'eron catalogue 2006 (\cite{Veron2006}). Radio Loud
objects are excluded from the sample according to the value of the
radio-loudness parameter ({\it R}; \cite{Stocke1992}). For QSO the condition
log(R) $>$ 1 is applied to reject a source, while in the case of Seyferts, in
addition to log(R) $>$ 2.4, sources are also excluded if log(R$_{\rm X}$) $>$ -2.755
(\cite{Panessa2007}), where R$_{\rm X}$ is the X-ray radio-loudness parameter
(\cite{Terashima2003}) (see \cite{Bianchi2009a} for details). No redshift
restriction has been imposed in our sample, which contains mostly local
sources with 90$\%$ of the sources at a redshift $\leq$ 0.5 and 60$\%$ at a
redshift $\leq$ 0.1. A distinction has been made between Broad Line (BL;
40$\%$ of full sample) and Narrow Line (NL; 20$\%$ of full sample) sources,
using as threshold the value of the Full Width Half Maximum (FWHM) of the
H$_{\rm \beta}$ line whenever available (65$\%$ of the sources in our sample):
H$_{\rm \beta}$ $\ge$ 2000~km/sec for BL and H$_{\rm \beta}$ $<$ 2000~km/sec for NL
(\cite{Osterbrock1987}). This distinction between NL and BL refers only to
this standard limit on H$_{\rm \beta}$, while no optical Type 2 objects are present
in the sample. When needed, and not derived in this work, source
properties are extracted from the CAIXA catalogue.

\subsection{The Flux-limited Sample}\label{FLSample}
\par The selection criteria with which CAIXA and FERO were assembled do not
provide a complete sample in the sense that the sources are not selected
according to a physical property: the main point of the FERO project is to
include as many objects as possible and this of course depends predominantly
on the availability of public observations. Because of the nature of the FERO
project, one of the fundamental requirements was to identify a complete,
unbiased subsample of sources with high signal to noise in order to derive
meaningful constraints on the properties of the (unknown) parent population of
local radio-quiet AGN. Hence, sources from the RXTE all-sky Slew Survey (XSS,
\cite{Revnivtsev2004}) having a count rate in the 3-8~keV energy band greater
than 1 cts/sec and fulfilling the FERO source selection criteria were
identified. This defines a flux-limited sample of 33 sources. The XSS is
nearly 80\% complete at the selected flux level for sources with Galactic
latitude greater than 10$^\circ$. For two of them, UGC~10683 and
ESO~0141-G055, no {\it XMM-Newton} data were available as of April 2008. This
leaves the number of XSS-selected bright sources in the FERO sample to 31
(listed in Table~\ref{tab_fls_src} in Appendix~\ref{appendixa}). Throughout
the paper, these 31 sources will be referred to as the flux-limited sample.

\section{Analysis}\label{Analysis}

\par The data corresponding to the 149 observations have been uniformly
analysed using SASv6.5 (\cite{Gabriel2004}) and the latest calibration files
available. Event lists were obtained for the EPIC-pn camera following standard
SAS data reduction procedures. Due to its larger effective area, only EPIC-pn
(\cite{Struder2001}) data has been considered in this work. The filtering of
event lists for periods of high background activity was performed by
maximizing the source signal to noise as in \cite{Piconcelli2004}. The
typical source extraction regions are circular and of the order of 50\arcsec
in radii for small window mode, while for full frame and large window modes,
the source extraction regions range from 20 to 40\arcsec in
radii. Background spectra were extracted from circular source-free regions,
except for observations taken in small window mode where blank-field event
lists were used as described in \cite{Read2003}, of a 50\arcsec in radii and
close to the target. Spectra affected by a pileup larger than 1$\%$ were
rejected. Full details of the data reduction and spectra accumulation are
reported in Bianchi et al. (2009a).

\subsection{XMM-Newton observations}\label{Observations}
\par The data of the observations presented here were public as of April
2008. All the observations are target observations with exposure times ranging
between 1~ksec and 400~ksec, with 90\% of the observations below 100~ksec.

\subsubsection{Multiple observations}\label{MultipleObservations}
\par When multiple observations of the same source were available, the
individual spectra were combined and treated, for all purposes, as a single
observation. The FTOOLS tasks {\it mathpha}, {\it adrmf}, and {\it addarf}
were used correspondingly to sum the source and background counts spectra and
the response matrix and ancillary files, using the exposure times of the
individual observations to be summed as weights in these last two
cases. However, spectra were only combined if the observations were taken with
the same observing mode and if the source was in {\it similar} flux state,
i.e., if the total flux and power-law spectral index in the 2-10 keV energy
band were consistent within the statistical errors. If these criteria were not
met, then the observation with the longest exposure time was chosen to prevent
introducing source-related bias in the selection. This procedure ensures that
mixing different spectral states from a given source does not affect the
results derived from our data set.

\par A total of 22 sources in our sample have multiple
observations. Table~\ref{mult_observations} lists on a source-by-source basis
{\it XMM-Newton} observation IDs that have been combined. For completeness,
Table~\ref{mult_observations_sing} in Appendix~\ref{appendixb} lists those
sources for which multiple observations are available, but only one was
considered; so no spectra were combined. This approach is different from
the one taken in CAIXA, where for sources with multiple observations, the one
with the longest exposure was selected systematically as independent of the
flux state. The importance of achieving high signal-to-noise spectra for the
FERO project, meant that it was preferred to co-add multiple observations of
the same source, when available.

\begin{table*}
\caption{List of sources within the FERO sample where multiple observations are
available.}
\label{mult_observations}
\centering          
\begin{tabular}{l c c c || l c c c }     
\hline\hline                             
\multicolumn{1}{c}{Source} & List of     & Observation & Observation & \multicolumn{1}{c}{Source}&  List of & Observation & Observation \\
\multicolumn{1}{c}{Name}   & {\it XMM-Newton}   & Date    & Exposure & \multicolumn{1}{c}{Name}  &   {\it XMM-Newton}&Date& Exposure \\ 
       & Observation &      & Time    &        &   Observation & & Time \\ 
       & IDs combined&      &    &        &   IDs combined& & \\ 
       &             &      &(ksec) &     &               & & (ksec)\\
\hline 
   PG1440+356 &{\bf 0005010101}& 2003-01-01&  17.2&          1H0419-577 &{\bf 0148000401}&2003-03-30 &11.1 \\ 
              &{\bf 0005010201}& 2003-01-04&  10.6&                     &{\bf 0148000501}&2003-06-25 &10.7 \\ 
              &{\bf 0005010301}& 2003-01-07&  18.1&                     &{\bf 0148000601}&2003-09-16 &11.3 \\ 
              &     0107660201 & 2001-12-23&  24.2&                     &     0148000201 &2002-09-25 &11.8 \\ 
       AKN564 &{\bf 0006810101}& 2000-06-17&   7.4&                     &     0148000301 &2002-12-27 & 7.6 \\ 
              &{\bf 0006810301}& 2001-06-09&   7.5&                     &     0148000701 &2003-11-15 & 1.1 \\ 
              &{\bf 0206400101}& 2005-01-05&  69.2&                     &     0112600401 &2000-12-04 & 5.7 \\ 
  MCG-6-30-15 &{\bf 0029740101}& 2001-07-31&  55.8&          PG1115+080 &{\bf 0203560201}&2004-06-10 &65.5 \\ 
              &{\bf 0029740701}& 2001-08-02&  85.7&                     &{\bf 0203560401}&2004-06-26 &71.2 \\ 
              &{\bf 0029740801}& 2001-08-04&  86.8&                     &{\bf 0082340101}&2001-11-25 &53.8 \\ 
              &{\bf 0111570101}& 2000-07-11&  28.8&      IRAS17020+4544 &{\bf 0206860101}&2004-08-30 &13.4 \\ 
              &{\bf 0111570201}& 2000-07-11&  37.9&                     &{\bf 0206860201}&2004-09-05 &12.3 \\  
      NGC5548 &{\bf 0089960301}& 2001-07-09&  58.7&              MRK507 &{\bf 0300910401}&2005-06-17 &15.8 \\ 
              &{\bf 0109960101}& 2000-12-24&  16.0&                     &{\bf 0300910701}&2005-09-01 &11.4 \\ 
              &     0089960401 & 2001-07-12&  19.8&              MRK279 &{\bf 0302480401}&2005-11-15 &41.4 \\ 
       MRK876 &{\bf 0102040601}& 2001-04-13&   2.6&                     &{\bf 0302480501}&2005-11-17 &40.7 \\ 
              &{\bf 0102041301}& 2001-08-29&   2.4&                     &{\bf 0302480601}&2005-11-19 &22.8 \\ 
  ESO15-IG011 &{\bf 0103861701}& 2000-09-29&   5.0&              MRK766 &{\bf 0109141301}&2001-05-20 &89.6\\
              &{\bf 0103862001}& 2001-10-31&   4.5&                     &{\bf 0304030301}&2005-05-25 &68.9\\
MCG-01-13-025 &{\bf 0103863001}& 2002-08-28&   4.3&                     &{\bf 0304030401}&2005-05-27 &65.8\\
              &{\bf 0103861401}& 2000-08-30&   1.5&                     &{\bf 0304030501}&2005-05-29 &64.9\\                      
     NGC 3516 &{\bf 0107460601}& 2001-04-10&  64.3&                     &{\bf 0304030601}&2005-05-31 &63.1\\                     
              &{\bf 0107460701}& 2001-11-09&  88.7&                     &{\bf 0304030701}&2005-06-03 &20.4\\
  REJ2248-511 &{\bf 0109070401}& 2000-10-26&  10.1&	                &     0096020101&2000-05-20 &25.7\\
              &{\bf 0109070601}& 2001-10-31&   9.8&			&     0304030101&2005-05-23 &66.2\\
    H0557-385 &{\bf 0109130501}& 2002-04-03&   4.0&SDSSJ135724.51+652505.9 &{\bf 0305920301}&2005-04-04 &21.1 \\ 	                 
              &{\bf 0109131001}& 2002-09-17&   6.5&                        &{\bf 0305920601}&2005-06-23 &12.0 \\ 	   
      TONS180 &{\bf 0110890401}& 2000-12-14&  20.3&       LBQS1228+1116 &{\bf 0306630101}&2005-12-13 &60.4 \\           
              &{\bf 0110890701}& 2002-06-30&  12.6&                     &{\bf 0306630201}&2005-12-17 &83.2 \\ 	   
      NGC7469 &{\bf 0112170101}& 2000-12-26&  12.3&              MRK509 &{\bf 0306090201}&2005-10-18 &59.8\\   	   
              &{\bf 0112170301}& 2000-12-26&  16.1&                     &{\bf 0306090301}&2005-10-20 &32.4\\      	   
              &{\bf 0207090101}& 2004-11-30&  59.2&                     &{\bf 0306090401}&2006-04-25 &48.6 \\ 	   
              &{\bf 0207090201}& 2004-12-03&  55.0&                     &{\bf 0130720201}&2001-04-20 &27.8 \\ 
      NGC3783 &{\bf 0112210101}& 2000-12-28&  26.1&                     &     0130720101 &2000-10-25 &20.7 \\ 
              &{\bf 0112210201}& 2001-12-17&  51.2&                     &                & &\\      	   
              &{\bf 0112210501}& 2001-12-19&  93.2&                     &                & &\\		   
\hline                  			                        		   
\end{tabular} 
\tablefoot{
Multiple observations were combined where available and treated,
for all purposes, as a single observation. Observation IDs in bold correspond
to those observations that were combined, while observation IDs not
marked in bold were discarded.}					                                        
\end{table*}
              
\subsection{Spectral analysis}\label{SpectralAnalysis}
\par The time-averaged spectra were re-binned in order not to oversample the
intrinsic energy resolution of the EPIC-pn camera ($\simeq$150~eV at 6~keV) by
a factor larger than 3, while making sure that each spectral channel contains
at least 25 background-subtracted counts to ensure the applicability of the
$\chi^2$ goodness-of-fit test. Fits were performed in the 2-10~keV energy
range. To ensure good quality spectral fits, only those source spectra with at
least 17 d.o.f. in this energy range were considered in the analysis. (12 out
of the 161 sources in the initial sample did not meet this criterion.) For the
spectral analysis and fitting, XSPEC v12.3.0 (\cite{Arnaud1996}) was
used. Throughout the analysis, solar abundances were assumed after
Anders \& Ebihara (1982).

\subsection{Spectral model}\label{SpectralModel}
\par 
All spectra were fitted with the following baseline model:

\begin{center}
\begin{equation}
e^{-\sigma N_{\rm H}} \cdot W(\Gamma,N^i_{\rm H},\xi) \cdot A [ E^{-\Gamma} +
C(\Gamma,R) + \sum_{i=1}^5 G_i + Ky(\theta,\beta,a)] \label{baseline}
\end{equation}
\end{center}

\par
where the different model components are
\begin{itemize}
\item[$\bullet$]{Galactic absorption (e$^{-\sigma N_{\rm H}}$), where $N_{\rm H}$ was fixed to the
galactic column density and $\sigma$ is the photoelectric cross section of
the process according to \cite{Morrison1983};}

\item[$\bullet$]{Power law (N(E) $\propto$ E$^{-\Gamma}$), where $\Gamma$ is
the photon index of the primary power-law X-ray source spectrum;}

\item[$\bullet$]{Compton reflection from neutral material ($C(\Gamma,R)$),
where {\it R} is the reflection factor as compared to material emitting over
2$\pi$ in solid angle (see however Murphy \& Yaqoob 2009 for some {\it
caveats} about this interpretation). The high-energy cut-off of the primary
spectrum was fixed to 100~keV. The viewing angle of the reflecting material
was fixed to 18$^\circ$, while the reflection normalization (R) was left free
during the fitting procedure. The model used was the XSPEC {\small PEXRAV}
component (\cite{Magdziarz1995});}

\item[$\bullet$]{Ionised absorption ($W(\Gamma,N^i_{\rm H},\xi)$) from warm
gas in the local AGN environment was introduced through the XSPEC model
{\small ABSORI}. Although more accurate and complex models exist for warm
absorption, they would overfit the lower-quality data in our sample when
uniformly applied to the whole sample. Both the intrinsic column density
($N^i_{\rm H}$) and ionisation parameter ($\xi$) were left free during the
fitting procedure, while the continuum slope ($\Gamma$) was tied to the photon
index of the primary power-law continuum. The temperature of the absorber was
fixed to 2$\times$10$^5$ K ($\sim$17~eV), representative of the temperature
found in warm absorbers (see e.g. \cite{Krongold2007});}

\item[$\bullet$]{Narrow lines (G$_{\rm i}$): 4 zero-width Gaussian lines were
included with centroid energies fixed at 6.4~keV (neutral Fe I K$_{\rm \alpha}$),
6.7~keV (ionised Fe XXV), 6.96~keV (ionised Fe XXVI), and 7.06~keV (neutral
Fe I K$_{\rm \beta}$). The normalization of the Fe I K$_{\rm \beta}$ line was limited to 
less than 16\% that of the Fe I K$_{\rm \alpha}$ line flux (\cite{Molendi2003}). The
two ionised lines were included in the baseline model as several studies
show that they represent an important spectral component in AGN. Among the
possible physical interpretations of these features are an origin in the
accretion disc (e.g. \cite{Reeves2001}, \cite{Pounds2001} and
\cite{Miller2007}) or produced in photoionised circumnuclear matter
(e.g. \cite{Bianchi2002} and \cite{Bianchi2005});}

\item[$\bullet$]{Fe I 6.4~keV Compton shoulder (G$_{\rm i}$). The Fe I K$_{\rm \alpha}$
Compton shoulder of the narrow component of the Fe I fluorescent line was
considered by using a Gaussian line at 6.3~keV, with 0.05~keV as the fixed width and
free normalization allowed to be no more than $\sim$20\% that of the
6.4~keV Fe I K$_{\rm \alpha}$ line, following \cite{Matt2002};}

\item[$\bullet$]{Relativistic line ($Ky(\theta,\beta,a$), {\it Kyrline}
model). To model the effect of the strong gravitational field on the emission
regions of the disc closest to the central black hole, the {\it Ky} set of
models within XSPEC was used (\cite{Dovciak2004}). The model considers
accretion disc emission around a rotating black hole with a given spin, {\it
a}. In contrast to models such as {\it diskline} or {\it laor}, {\it Kyrline}
allows the spin to be fitted as a free parameter. For the line emission, a
fixed centroid energy of 6.4~keV (in the rest frame of the source), is
assumed, while the emission between the innermost stable orbit and 400
gravitational radii is integrated, keeping these limits fixed during the
fitting procedure. The radial dependence of the disc emission is modelled with
a power law of index $\beta$ ($\propto$ r$^{-\beta}$), while Laor's limb
darkening law was adopted to characterize the angular dependence. Here, the
radial disc emissivity is referred to as $\beta$. The disc inclination angle
($\theta$), $\beta$, and {\it a} are free parameters during the fitting
procedure.}

\end{itemize}

\par In summary, the model that was uniformly fitted to our data consists of 9
components with a total of 14 free independently-fitted parameters. For those
cases in which the value of the FWHM of the H$_{\rm \beta}$ line is available,
the model has 15 free parameters (see Section~\ref{Test_Mod_Par}). The fitting
procedure was carried out over different steps. First, the power law with
neutral reflection and absorption were fitted to the data, followed by the
addition of the narrow line components one by one (in the order listed above).
The last model component added to the fit was the relativistic line. After each
addition of a new component, a fit was performed, and no model component
discarded regardless of the goodness-of-fit. This implies that the $\chi^2$
values obtained from the fits correspond to the best-fit given by this
model. Its important to stress that no attempt has been made as a whole or on
a source-by-source basis to find the best-fit model yielding the lowest
possible $\chi^2$ value. For reference, table \ref{tab_fls_model} gives, for
those sources belonging to the flux-limited sample, the most relevant
parameters of the best fit for the model considered.

\subsubsection{Limitations to model parameters}\label{Test_Mod_Par}

\par
During the fitting procedure, some limitations are introduced for the
following parameters of the baseline model:

\begin{itemize}
\item[$\circ$] The disc inclination angle in the {\it Kyrline} model was
  limited to a maximum value of 60$^\circ$. In the Seyfert unification
  scenario (\cite{Antonucci1993}), Type 1 objects are not expected to have
  large viewing angles. It has been found that relaxing the restriction on the
  inclination angle yields a number of detections of broad and highly skewed
  lines with equivalent widths of the order of 1~keV. This effect is simply
  because in sources with poor statistics above 7~keV, the broad line model
  fits the continuum, also reflected by required large disc-inclination angles
  ($>$80$^\circ$).
\item[$\circ$] The disc emissivity in the {\it Kyrline} model was limited
  to a maximum value of 6 given that steeper profiles are not expected, even
  taking strong general relativistic effects into account on the primary
  emission such as light bending (see \cite{Miniutti2004}).
\item[$\circ$] To possibly constrain the parameter space, limitations to the
  spin parameter in the {\it Kyrline} model were tested. Three fitting runs
  were produced with black hole spin fixed to 0.998 ({\it Kerr}), fixed to 0
  ({\it Schwarzschild}), and free to vary between 0 and 0.998. In the first
  two cases, no systematic difference was found when comparing the results in
  terms of $\chi^2$ improvement, leading us to conclude that none of the two
  extreme spin values can be assumed {\it a priori}. Therefore, the results
  presented are extracted from the run where the black hole spin is left free
  to vary during the fitting procedure.
\item[$\circ$] The power-law index ($\Gamma$) and the reflection fraction (R)
  were limited following the work by \cite{Dadina2008} on a {\it BeppoSAX}
  sample of Type 1 Seyfert galaxies. Given the distributions of $\Gamma$ and R
  derived by \cite{Dadina2008}, a limit of three times the 1$\sigma$ standard
  deviation of the average values is chosen for $\Gamma$ and R ($\Gamma \in$
  [0.45:3.33] ; R $\in$ [0.:3.33]).
\item[$\circ$] When available in the literature, the value of the FWHM of the
  optical H$_{\rm \beta}$ line was used to set an upper limit to the width of
  the narrow component of the 6.4~keV neutral Fe I K$_{\rm \alpha}$ line,
  leaving the width of the line free. This choice is consistent with the
  possibility that for some sources, the Fe I K$_{\rm \alpha}$ line may be
  produced in the optical broad line region (see e.g. \cite{Bianchi2008} and
  references therein). When the H$_{\rm \beta}$ FWHM was not available, the Fe
  I K$_{\rm \alpha}$ line width was assumed to be unresolved ($\sigma\equiv$
  0).
\end{itemize}

\subsubsection{Test of spectral model components}\label{Test_Mod_Comp}

\par During the process of selecting the baseline model, several tests were
performed with different model components. The most important ones are
summarised below.

\begin{itemize}
\item[$\circ$] To contemplate the hypothesis that the reprocessed AGN
  continuum in Type 1 objects is made by contribution of both torus and disc
  reflection, the presence of two reflection components at the same time was
  tested. The second {\small PEXRAV} disc component was relativistically
  blurred using the convolution model {\it kdblur}, to take possible
  relativistic effects into account. After close inspection of the
  contribution from each physical system, it was concluded that the
  sensitivity band of {\it XMM-Newton} does not allow disentangling the
  reflection component from the torus from that of the disc. Since the
  presence of torus reflection is very well supported observationally by the
  almost ubiquitous narrow Fe~I~K$_{\rm \alpha}$ component
  (\cite{Bianchi2004}; \cite{Yaqoob2004}), only this {\small PEXRAV} component
  has been included in our baseline model.
\item[$\circ$] The effect of reflection from an ionised accretion disc was
  tested by replacing the torus reflection with the model component {\small
  PEXRIV} (\cite{Magdziarz1995}): Eq.~\ref{baseline} is thus modified to
  include the disc ionisation parameter $\xi$ in the function C($\Gamma$, R),
  which is left free during the fitting procedure. The centroid energy in the
  {\it Kyrline} model was consistently set to 6.7~keV. The results from this
  test are discussed in Sects.~\ref{Results} and~\ref{Discussion}.
\end{itemize}

\section{Results}\label{Results}

\par The EW of the {\it Kyrline} component was used as a proxy for
establishing significant detections of relativistic Fe I K$_{\rm \alpha}$
lines in the sample. The line EW is defined in this work as the ratio of the
flux of the {\it Kyrline} component to the flux of the continuum, as defined
in Section~\ref{SpectralModel}, integrated over the energy range between
$\pm$5\% around the line rest-frame energy (6.4~keV). The error in the
normalization of the {\it Kyrline} component is used to estimate the
uncertainty in the corresponding line EW. The threshold for significant
detections is set to 5$\sigma$ confidence level (c.l.). This is a somewhat
more conservative approach than normally taken when reporting relativistic
line detections in individual objects, but it has to be kept in mind that the
FERO project is a statistical study of a large sample that includes spectra of
very different levels of statistical quality. For those sources with no
significant line detection, the upper limits to the line EW at the 90\%
c.l. are provided. In the following, errors, upper and lower limits, on the
relevant parameters are given at the 90$\%$ c.l. for one interesting parameter
($\Delta\chi^2$ = 2.71).

\subsection{Sources with detected broad Fe lines.}

\par
\begin{figure*}
  \centering
  \includegraphics[width=.8\textwidth]{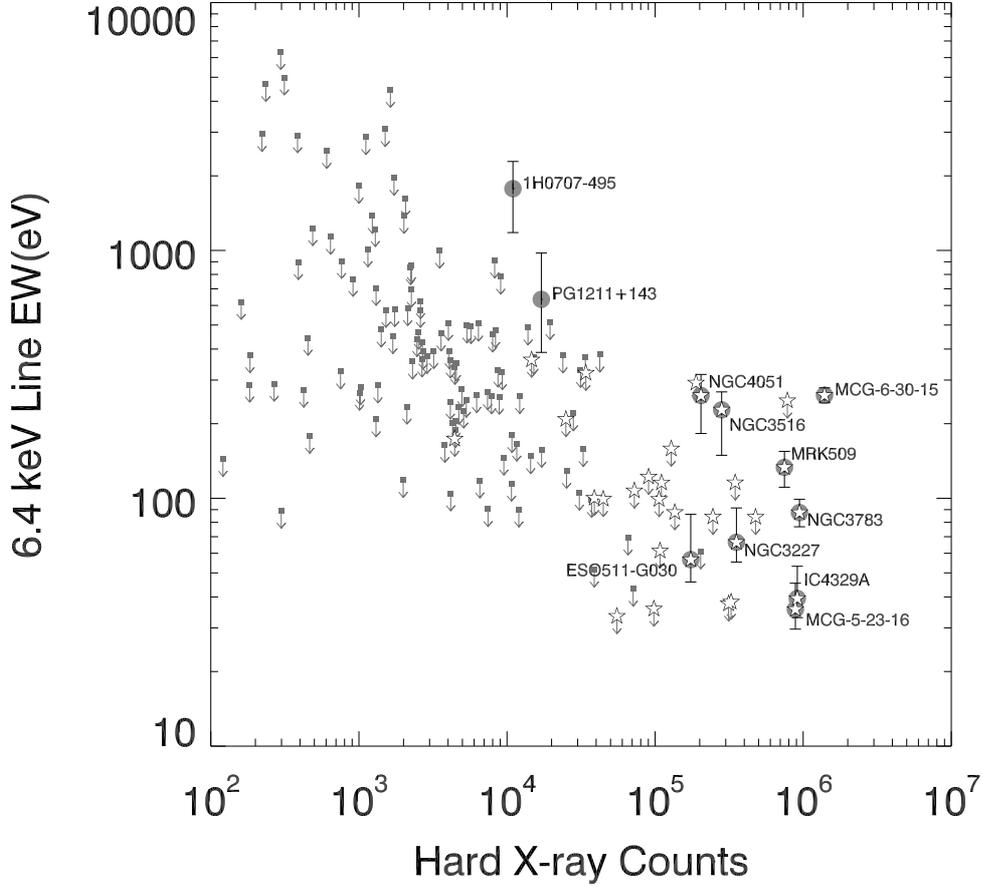}
  \caption{Equivalent width of the relativistic broad 6.4~keV Fe K${\rm _\alpha}$
  line vs. hard X-ray counts (2-10~keV) for the FERO sources. Filled circles
  indicate line detections at $\ge$5$\sigma$ confidence level (where error
  bars indicate the 90\% confidence level intervals), while filled squares
  indicate line upper limits at the 90\% confidence level. White stars
  indicate sources belonging to the flux-limited sample (see
  Section~\ref{FLSample} for details).}
  \label{fig_detections}
\end{figure*}

 Figure~\ref{fig_detections} shows the EW of the broad Fe K${\rm _\alpha}$
line at 6.4~keV as a function of source counts in the 2-10~keV band. This plot
shows two things. First, significant broad Fe K${\rm _\alpha}$ line detections
are concentrated in the portion of the plot dominated by spectra with good
statistical quality (roughly with $\gtrsim$ 1.5$\times$10$^5$ hard X-ray
counts). Second, overall, the distribution of line EW upper limits versus hard
X-ray counts follows the expected trend; i.e., in the absence of a significant
broad line, the sensitivity (the ability to detect a broad line) goes as
(1/$\sqrt{(cts_{\rm 2-10keV})}$.  The dispersion of EW upper limits around
this direction, for a given range of hard X-ray counts, could be explained in
terms of scattering of the model parameters.

\par Table~\ref{tab_detections} provides the detailed list of the significant
relativistic broad Fe K$_{\rm \alpha}$ line detections with the best-fit
parameters of the {\it Kyrline} model. From now on, this table is referred to
as the list of detected relativistic Fe K$_{\rm \alpha}$ lines in the FERO
sample. In the fitting procedure, relativistic effects from neutral and
ionised reflection were tested as explained in Section
\ref{Test_Mod_Comp}. The test for a neutral Fe K$_{\rm \alpha}$ broad line
yields in total 11 significant detections at $\ge$5$\sigma$ c.l.. The fit with
the model with an ionised Fe K$_{\rm \alpha}$ line yields a detection in all
the 11 sources, except IC4329A, 1H0707-495, and MCG-5-23-16, which only
provide detection in the neutral case. In the 8 sources where a relativistic
broad Fe K$_{\rm \alpha}$ line is equally detected by means of neutral and
ionised reflection models, the best-fit models were carefully examined and the
model with the lowest $\chi^2$ preferred. MRK~509 is the only source for which
the ionised reflection provides a significant better-fit statistic (with the
probability of the improvement being by chance of
2.2$\times$10$^{-7}$). MRK~766 and ARK~120 on the contrary, show significant
broad Fe lines only in the ionised case. At the end, neutral and ionised runs
were merged, which yields 13 sources with strong evidence of a broad Fe
line. Last, it is worth pointing out that the only source where a relativistic
broad Fe K$_{\rm \alpha}$ line is claimed for the first time in this work is
ESO511-G030, with an EW of 57$^{+30}_{-11}$~eV.

\par Two sources, PG1211+143 and 1H0707-495, are characterised by extremely
high EW of the Fe K$_{\rm \alpha}$ line (Figure~\ref{fig_detections}, see also
Table~\ref{tab_detections}). The data included in our sample for both sources
have similar spectral shapes, showing deep spectral features at $\sim$7~keV
and spectral curvature in the Fe energy band. Pounds et al. (2003a) argue that
PG1211+143 presents a high column density (5 $\times$ 10$^{23}$ cm$^{-2}$) of
highly ionised matter along the line of sight partially covering the central
hard X-ray source and that the apparent presence of a relativistic Fe K${\rm
_\alpha}$ emission line could be an artifact of absorption. However, the
relativistic broad line interpretation is not discarded, and the same authors
find that a broad line with a large EW, of the order of 600-900~eV, and a
combination of absorption features, provide a good interpretation of the
data. This value of the Fe K$_{\rm \alpha}$ line EW is in good agreement with
the value found in this work (635$^{+341}_{-247}$~eV). Nevertheless, whether
there is a relativistically broadened Fe K$_{\rm \alpha}$ line in this source
is debatable, as new results from a more recent analysis on data from 2007
indicate (\cite{Pounds2009}). 1H0707-495 also shows a prominent flux drop
around 7~keV, which can be modelled with an absorption edge assuming large Fe
overabundance (\cite{Gallo2004}). Fabian et al. (2004) interpret the drop
around 7~keV in terms of relativistically blurred ionised reflection from the
accretion disc, and derive an EW for the broad Fe K$_{\rm \alpha}$ line of
1800~eV, in agreement with the value found in the FERO analysis
(1775$^{+511}_{-594}$~eV). Fabian et al. (2004) proposed that the X-ray
spectrum of this source is disc reflection dominated, and as the high value of
the line EW implies, high element abundances were required, in a similar way
to absorption-dominated models (\cite{Gallo2004}). Recently, Fabian et
al. (2009) have presented new timing-based arguments in favour of the
relativistic nature of the broad features in this source with the simultaneous
detection of broad Fe L and K lines\footnote{We point out that the {\it XMM-Newton}
spectra of 1H0707-495 included in the FERO sample correspond to the ones used
by Fabian et al. (2004) and that Fabian et al. (2009) used a different
dataset.}. However, Miller et al. (2010) has recently claimed that the Fabian
et al. (2009) interpretation of the timing properties of the source is not
unique and that solutions that do not invoke relativistic reprocessing might
still be viable.

\begin{table*}
\caption{Sources within the FERO sample where a relativistic Fe K$_{\rm
\alpha}$ line is detected with a significance $\ge$5$\sigma$.}
\label{tab_detections}   
\centering          
\begin{tabular}{l l c c c c c c c l}     
 \hline\hline
Source & Type$^{(1)}$ & L$_{\rm X}^{\rm 2-10~keV}$ & Cts$^{\rm 2-10~keV}$&
\multicolumn{1}{c}{EW} & $\theta$ & a & $\beta$ & $\chi^2$/dof & References$^{(2)}$\\

&       &  (10$^{42}$ erg s$^{-1}$) & (10$^5$ cts) & \multicolumn{1}{c}{(eV)}  
& ($^\circ$) &         &  & & \\ 
\hline                    
IC4329A & BLSY & 56.4 & 9.123$\pm$0.009 &  39$^{+14}_{-6}$&    28$^{+6}_{-11}$ & $\ge$0.0		      &  $<$1.3 & 238.1/161 &  {\bf (1)}, {\bf (2)}, (3), (4)\\  
MCG-5-23-16 & NCSY & 14.2 & 8.853$\pm$0.009 &  36$^{+10}_{-6}$& 21$^{+8}_{-3}$ & $\ge$0.0 		      &  $<$1.6 & 271.4/162 &  {\bf (5)}, {\bf (6)}\\ 
ESO511-G030 & NCSY & 22.4  & 1.740$\pm$0.004 & 57$^{+30}_{-11}$& 18$^{+7}_{-7}$ & $\ge$0.0 &  $<$1.1 & 198.2/161  & \\ 
MCG-6-30-15 & BLSY &  5.4 &13.927$\pm$0.012 & 260$^{+19}_{-18}$&    40$^{+1}_{-3}$ & 0.86$^{+0.01}_{-0.02}$ &  4.1$^{+0.2}_{-0.2}$ & 227.6/161  &  {\bf (7)}, {\bf (8)}, {\bf (9)}, (10)\\ 
NGC4051 & NLSY &  0.3 & 2.040$\pm$0.004 & 260$^{+56}_{-77}$&    22$^{+6}_{-6}$ & $>$0.46		      &  2.9$^{+0.3}_{-0.4}$ & 198.8/162 & {\bf (11)}, {\bf (12)}, (13)\\ 
NGC3516 & BLSY &  3.6 & 2.810$\pm$0.005 & 227$^{+41}_{-78}$&    27$^{+2}_{-3}$ & $>$0.48	              &  2.8$^{+0.2}_{-0.3}$ & 223.1/161 &  {\bf (14)}, {\bf (15)}, (16)\\ 
NGC3783 & BLSY & 11.4 & 9.440$\pm$0.010 &  88$^{+11}_{-11}$&  $<$8    	       & $>$0.16		      &  2.7$^{+0.1}_{-0.2}$ & 304.5/161 &  {\bf (17)}, (18)\\ 
NGC3227 & BLSY &  1.2 & 3.541$\pm$0.006 &  66$^{+25}_{-11}$&    23$^{+4}_{-4}$ & $\ge$0.0 &  1.9$^{+0.6}_{-0.5}$ & 230.9/162 & {\bf (19)}, (20)\\
MRK509  (*)  &  BLQ &103.5 & 7.471$\pm$0.009 & 170$^{+26}_{-19}$& 53$^{+1}_{-1}$& 0.78$^{+0.03}_{-0.04}$& $>3.8$ & 277.3/158 &  {\bf (21)}, {\bf (22)}\\ 
MRK766  (*)  & NLSY &  6.1 & 7.816$\pm$0.008 & 234$^{+23}_{-49}$& 20$^{+3}_{-2}$& $>$0.47& 2.7$^{+0.2}_{-0.1}$ & 300.3/160 &  {\bf (23)}, (24)\\
ARK120  (*)  & BLSY & 90.5 & 3.473$\pm$0.006 & 140$^{+31}_{-41}$& $>$59& $\ge$0.0 & 2.2$^{+0.6}_{-0.3}$ & 212.3/158 &  {\bf (25)}\\
\\
1H0707-495 & NLSY & 46.8  & 0.110$\pm$0.001 & 1775$^{+511}_{-594}$& 54$^{+3}_{-3}$ & $\ge$0.93 & 4.13$^{+1.22}_{-0.65}$& 107.6/116 & {\bf (26)},{\bf (27)},(28) \\
PG1211+143 & NLQ  & 49.4  & 0.171$\pm$0.001 &  635$^{+341}_{-247}$& 52$^{+4}_{-7}$ & $\ge$0.92 & $>$4.37 &  160.7/139 & {\bf (29)}, (30) \\
\hline \\
\multicolumn{4}{l}{(*) 6.7~keV K$_{\rm \alpha}$ line is preferred}\\
\end{tabular}
\tablefoot{* Sources for which there is a significant detection of a
relativistic {\it ionised 6.7~keV K$_{\rm \alpha}$ line} (see text for
details). The first 11 sources listed in the table belong to the flux-limited
sample. $\theta$, {\it a}, and $\beta$ are the disc inclination angle with
respect to the observer, the black hole spin and disc emissivity, respectively,
of the best-fit parameters of the {\it Kyrline} model. Except for the hard
X-ray counts where 1$\sigma$ errors are given, errors in any given parameter
are given at the 90\% confidence level. Also, upper and lower limits in any
relevant paremeter are given at the 90\% confidence level.\\
\tablefoottext{1}{SY: Seyfert, Q: quasar, NL: narrow line, BL: broad line, NC:
not classified.}  \tablefoottext{2}{References where evidence for a
relativistic Fe K${\rm _\alpha}$ line for a particular object is reported
(references marked in bold) as well as those where alternative scenarios are
provided.  The list of references is not meant to be complete.}  } \tablebib{
(1)~\cite{Done2000}; (2)~\cite{McKernan2004}; (3)~\cite{Steenbrugge2005};
(4)~\cite{Gondoin2001}; (5)~\cite{Reeves2007}; (6)~\cite{Braito2007};
(7)~\cite{Wilms2001}; (8)~\cite{Fabian2002}; (9)~\cite{Vaughan2004b};
(10)~\cite{Miller2008}; (11)~\cite{Ponti2006}; (12)~\cite{Uttley2004};
(13)~\cite{Pounds2004}; (14)~\cite{Iwasawa2004}; (15)~\cite{Nandra1999};
(16)~\cite{Turner2008}; (17)~\cite{Tombesi2007}; (18)~\cite{Reeves2004};
(19)~\cite{Markowitz2009}; (20)~\cite{Gondoin2003}; (21)~\cite{Ponti2009};
(22)~\cite{Page2003}; (23)~\cite{Pounds2003b}; (24)~\cite{Miller2007};
(25)~\cite{Vaughan2004a}; (26)~\cite{Fabian2009}; (27)~\cite{Fabian2004};
(28)~\cite{Gallo2004}; (29)~\cite{Pounds2003a}; (30)~\cite{Pounds2009}.  }
\end{table*}

\par
\begin{figure*}
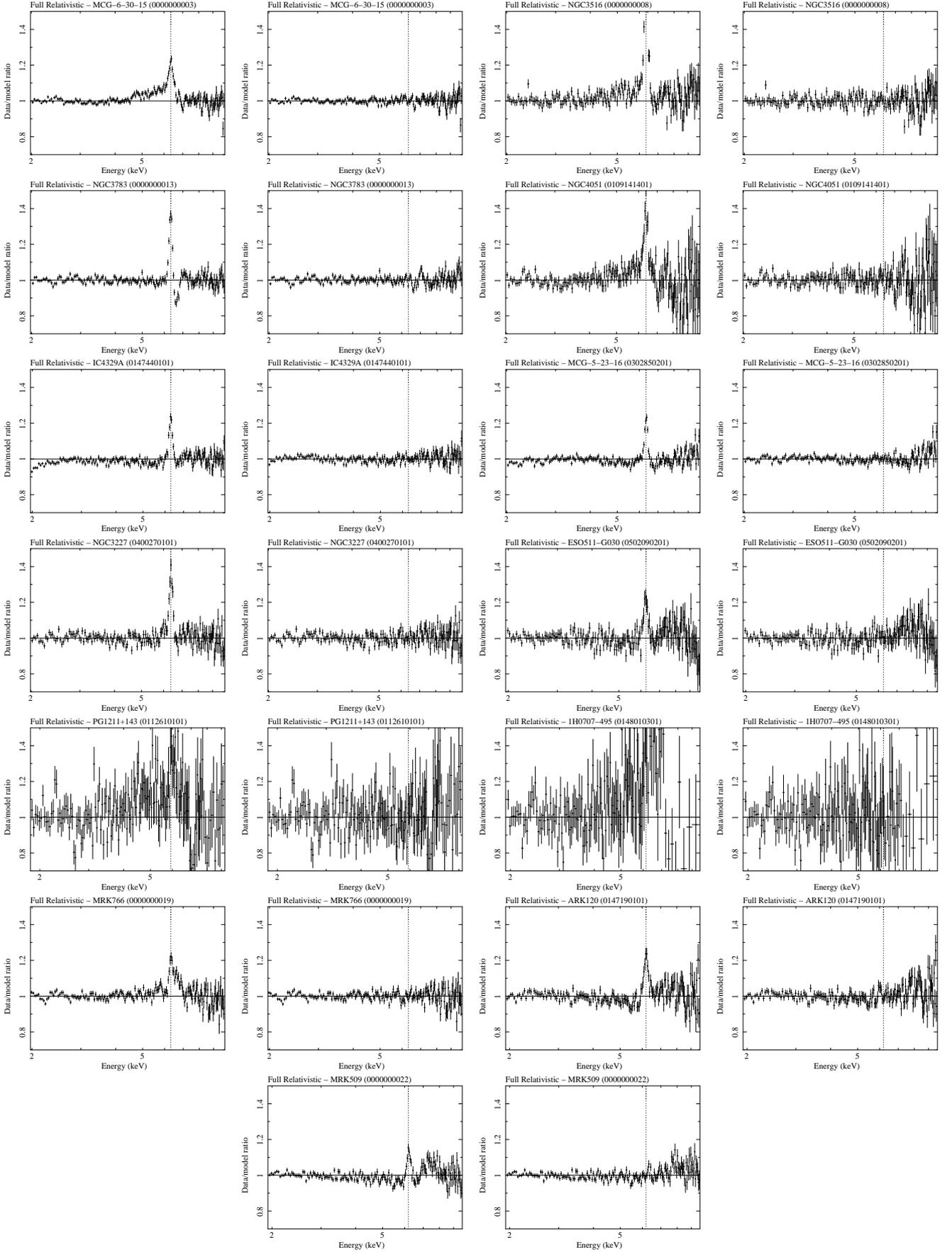

  \centering 
  \includegraphics[width=.18\textwidth,angle=270]{0000000003_nice_ratio_plot_v1.ps}
  \includegraphics[width=.18\textwidth,angle=270]{0000000003_nice_ratio_plot_rel_v1.ps}
  \includegraphics[width=.18\textwidth,angle=270]{0000000008_nice_ratio_plot_v1.ps}
  \includegraphics[width=.18\textwidth,angle=270]{0000000008_nice_ratio_plot_rel_v1.ps}\\
  \includegraphics[width=.18\textwidth,angle=270]{0000000013_nice_ratio_plot_v1.ps}
  \includegraphics[width=.18\textwidth,angle=270]{0000000013_nice_ratio_plot_rel_v1.ps}
  \includegraphics[width=.18\textwidth,angle=270]{0109141401_nice_ratio_plot_v1.ps}
  \includegraphics[width=.18\textwidth,angle=270]{0109141401_nice_ratio_plot_rel_v1.ps}\\
  \includegraphics[width=.18\textwidth,angle=270]{0147440101_nice_ratio_plot_v1.ps}
  \includegraphics[width=.18\textwidth,angle=270]{0147440101_nice_ratio_plot_rel_v1.ps}
  \includegraphics[width=.18\textwidth,angle=270]{0302850201_nice_ratio_plot_v1.ps}
  \includegraphics[width=.18\textwidth,angle=270]{0302850201_nice_ratio_plot_rel_v1.ps}\\
  \includegraphics[width=.18\textwidth,angle=270]{0400270101_nice_ratio_plot_v1.ps}
  \includegraphics[width=.18\textwidth,angle=270]{0400270101_nice_ratio_plot_rel_v1.ps}
  \includegraphics[width=.18\textwidth,angle=270]{0502090201_nice_ratio_plot_v1.ps}
  \includegraphics[width=.18\textwidth,angle=270]{0502090201_nice_ratio_plot_rel_v1.ps}\\
  \includegraphics[width=.18\textwidth,angle=270]{0112610101_nice_ratio_plot_v1.ps}
  \includegraphics[width=.18\textwidth,angle=270]{0112610101_nice_ratio_plot_rel_v1.ps}
  \includegraphics[width=.18\textwidth,angle=270]{0148010301_nice_ratio_plot_v1.ps}
  \includegraphics[width=.18\textwidth,angle=270]{0148010301_nice_ratio_plot_rel_v1.ps}\\
  \includegraphics[width=.18\textwidth,angle=270]{0000000019_nice_ratio_plot_v1.ps}
  \includegraphics[width=.18\textwidth,angle=270]{0000000019_nice_ratio_plot_rel_v1.ps}
  \includegraphics[width=.18\textwidth,angle=270]{0147190101_nice_ratio_plot_v1.ps}
  \includegraphics[width=.18\textwidth,angle=270]{0147190101_nice_ratio_plot_rel_v1.ps}\\ 
  \includegraphics[width=.18\textwidth,angle=270]{0000000022_nice_ratio_plot_v1.ps}
  \includegraphics[width=.18\textwidth,angle=270]{0000000022_nice_ratio_plot_rel_v1.ps}\\

  \caption{Set of two figures for those sources with a relativistic 6.4~keV
  (6.7~keV) Fe K$_{\rm \alpha}$ line EW detection $\ge$5$\sigma$ confidence
  level: data to the best-fit continuum ratio (left set column), and data to
  the best-fit model ratio (right set column). The vertical dashed line is
  placed for reference at 6.4~keV (6.7~keV) in the rest frame of the
  source. On top of each figure the model used for the ratio, the
  corresponding source name, and the observation ID are given. For those
  sources with summed observations a {\it dummy} observation ID is given.}
  \label{fig_ratios_det1}
\end{figure*}

 Figure~\ref{fig_ratios_det1} shows the ratios of the data to the
continuum and data to the best-fit model in the 2-10 keV energy range for the
sources with a relativistic 6.4~keV Fe K$_{\rm \alpha}$ broad line
detection. For MRK509, MRK766 and ARK120, the figures were produced with the
model including the ionised continuum and 6.7~keV Fe K$_{\rm \alpha}$
line. The latter set of plots indicates how well the baseline model used in
this work describes the data. The deviation of the best-fit model and the data
is limited below $\sim$10\% in the whole 2-10 keV energy range used for
fitting, and below $\sim$5\% if one excludes the last energy bins ($>$8~keV)
with poorer statistics. Despite the fact that by selection the sources in FERO
are unobscured ($N_{\rm H}<2\times10^{22}$cm$^{-2}$) for two sources,
MCG-5-23-16 and IC4329A, a downturn in the ratio plot below 3~keV is
noticeable and could indicate extra cold absorption, which in principle would
not have been taken into account by our baseline model. However, for the
particular case of these two sources, it is believed that this downturn cannot
be attributed to extra cold absorption since the parameters derived from the
warm absorption model ({\small ABSORI}) used yield values compatible with
those extracted from CAIXA (as seen in Table~\ref{tab_fls_model}), $N_{\rm H}$
(z=0) (1.20$^{+0.13}_{-0.02}$) $\times$ 10$^{22}$~cm$^{-2}$, and
(0.33$^{+0.04}_{-0.02}$) $\times$ 10$^{22}$~cm$^{-2}$ for MCG-5-23-16 and
IC4329A, respectively, and in both cases, compatible with neutral absorption
($\xi<$ 0.02~erg~cm$^{-1}$ s$^{-1}$ for MCG-5-23-16 and $\xi<$
0.47~erg~cm$^{-1}$~s$^{-1}$ for IC4329A). For these two
sources, any extra neutral absorption that is present should have been taken
into account by the model component {\small
ABSORI}. Table~\ref{tab_detections} lists in the last column the $\chi^2$ and
corresponding number of degrees of freedom for those sources with 
evidence of a relativistic Fe K$_{\rm \alpha}$ line.

\subsection{Fraction of relativistic Fe lines}\label{fraction_det}

\subsubsection{Detection fraction in FERO}\label{fraction_det1}

\par The fraction of relativistic Fe lines detected in the FERO sample is 9\%
(13/149). Considering only the sources in the flux-limited sample, the
detection fraction rises to 36\% (11/31). These two numbers do not provide
specific information on the {\it intrinsic} fraction of AGN that have broad Fe
lines. They are limited by FERO being made of spectra of disparate quality and
by the unavailability of a well-defined complete AGN sample. Nevertheless, the
observed detection fraction can be considered as a lower limit for the
intrinsic number of AGN that would show a broad Fe line if, for example, all
sources were observed with the same signal-to-noise.

\par It is possible to estimate statistical errors on the detection fractions
to reflect the uncertainty introduced by considering a limited number of
sources. Detection fractions are calculated as {\it val$_1$/val$_2$}, where
{\it val$_1$} and {\it val$_2$} indicate a given number of sources used to
work out a given detection fraction. For the errors of {\it val$_{1,2}$},
Poisson statistics are assumed, where the errors are calculated as the average
of the upper ($\Delta_{\rm UL}$) and lower limit ($\Delta_{\rm LL}$) of the 84\%
c.l. (which correspond to the commonly used $\Delta_{\rm UL}$ = 1 + $\sqrt{{\it
val_{1,2}} + 0.75}$ and $\Delta_{\rm LL}$ = $\sqrt{{\it val_{1,2}} - 0.25}$,
according to Gehrels (1986)), both of which are good approximations for this
moderate confidence level and are accurate to better than 5\% even for low
values of {\it val$_{1,2}$} (see Gehrels 1986 for more information). The
error of {\it val$_1$/val$_2$} is propagated quadratically. In this way, the
detections fractions and their associated errors are 9$\pm$3\% and 36$\pm$14\%
for the FERO sample and for the flux-limited sample, respectively.

\par A broader understanding of the incidence of the relativistic line in
our work can be achieved by considering the upper limits on the line intensity
that characterize the majority of the sources. Within the FERO sample,
$\sim$16$\pm$4\% of broad line EW upper limits are below 100~eV. This value
corresponds to the EW of a neutral disc with an inclination of 60$^\circ$, for
an illuminating power law of $\Gamma$=2 and solar abundances
(\cite{Matt1992}). If the analysis is restricted to the 20 sources with no
significant broad line within the flux-limited sample, the fraction of sources
in which the line EW is limited below 100~eV is 50\% (10/20) (for reference,
for 75, 50 and 25~eV, the fractions are respectively 25\%, 20\%,
and $<$13\%). However, 5 sources within the flux-limited sample have measured
EWs lower than and inconsistent with this threshold, implying deviations from
the standard scenario either in terms of very high disc inclination (unlikely
given the relatively unobscured nature of these AGN), element abundances lower
than solar or a non-standard accretion disc geometry (\cite{George1991}).

\par Last, Table~\ref{FL_UL} in Appendix~\ref{appendixd} lists the 20 sources
with broad line upper limits in the flux-limited sample. In particular, four
sources (MRK~279, NGC~5548, ESO~198-G24 and MRK~590) exhibit upper limits
below 40~eV, arbitrarily chosen to be comparable to the lowest line EW
detected. Explaining line EWs below this value in the framework of standard
accretion theories is not straightforward, and it requires the use of
non-standard system properties. Assuming the above threshold, the {\it
non-detection fraction} of broad lines in the flux-limited sample can then be
calculated as 13\% (4/31). Otherwise stated, the presence of a broad line
cannot be excluded {\it a priori} in the remaining 87\% of the sources in the
flux-limited sample (although the line is formally detected only in 36\% of
the sources at a 5$\sigma$ level). This fraction is interpreted here as the
upper limit to the fraction of AGN exhibiting relativistically broadened Fe
lines in the parent population.

\subsubsection{Detection fraction and warm absorber}

\par In some cases, the use of the model component {\small ABSORI} could be
regarded as too simplistic, however necessary for the approach considered
here. Nevertheless, the influence of the selected warm absorption component
for the baseline model in determining the detection fraction was tested by
running a consistency check over the 31 sources belonging to the flux-limited
sample. The test replaces the model component {\small ABSORI} by the more
complex and complete {\small ZXIPCF} model and looks for a significant change
in the fraction of detected broad lines and the average broad line EW. {\small
ZXIPCF} uses a grid of XSTAR photoionised absorption models and allows
defining a covering fraction, which has been fixed to 1, and column density
and ionisation parameters of the absorbing material, which have been left free
during the fitting procedure. The results of this test show that the
introduction of the {\small ZXIPCF} model component does not significantly
provide a better fit, and yield a detection fraction of broad iron K$_\alpha$
lines of 39$\pm$15\% (12/31). This result is fully consistent with the
36$\pm$14\% (11/31) derived when using {\small ABSORI} for the warm
absorption, and concludes that the detection fraction is not significantly
affected by the approach taken in this work to model the warm absorber. Out of
the sources from the flux-limited sample listed in Table~\ref{tab_detections},
only ESO511-G030 does not show evidence of a broad line at the 5
sigma level when using {\small ZXIPCF}. In turn, MRK704 and NGC4593 show
a broad line at the 5.5$\sigma$ and 5.3$\sigma$ levels,
respectively, both at the limit of our detection threshold of 5~$\sigma$. The
average line EW derived from the detected lines is $<$EW$>$=154$\pm$112~eV
($<$EW$>_w$=73$\pm$3~eV), again both consistent with the results derived when
using the absorption model {\small ABSORI} (see
Section~\ref{LineDetectionsEW}).

\subsubsection{Detection fraction and X-ray luminosity}\label{fraction_det2}

\par Eleven out of the 13 lines detected correspond to sources classified as
Seyfert galaxies, while only 2 come from a source classified as quasar. In a
similar manner to the previous section, this corresponds to a
detection fraction of 13$\pm$5\% and $\leq$6\% for Seyfert galaxies and
quasars respectively. If the exercise is restricted to sources in the
flux-limited sample, the detection fractions are 36$\pm$15\% and $\leq$92\%
for Seyfert galaxies and quasars, respectively. Owing to the low number of lines
detected in quasars and the low number of quasars present in the flux-limited
sample, only upper limits to the detection fraction can be derived.

\par
\begin{figure}
  \centering
  \includegraphics[width=.45\textwidth]{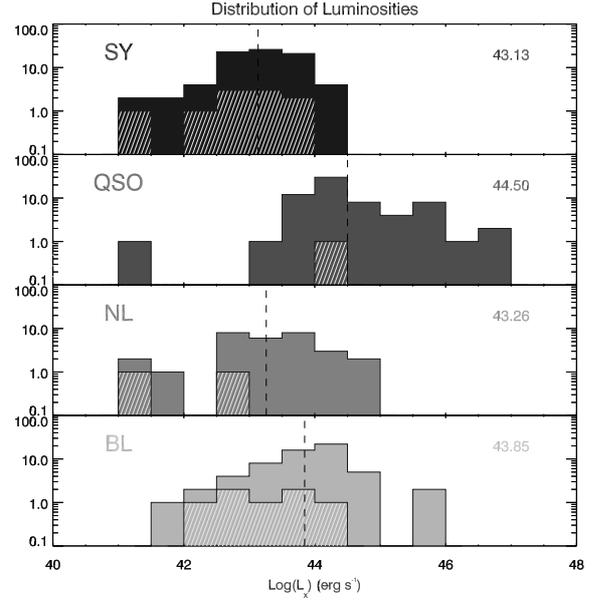}
  \caption{X-ray luminosity distribution in the 2-10~keV energy band for the
  FERO sample. This sample has been split into Seyfert galaxies (Sy; 82
  sources), quasars (QSO; 67 sources), narrow line (NL; 30 sources) and broad
  line (BL; 60 sources). The shaded area corresponds to the luminosity of
  those sources within the flux-limited sample for which a significant
  relativistic broad Fe K$_{\rm \alpha}$ line detection has been found. The
  numbers shown on the right in each panel are the median values of the
  correspondent distribution, also marked by the vertical dashed line.}
  \label{fig_luminosity}
\end{figure}

 Figure~\ref{fig_luminosity} shows the distribution of the 2-10~keV
luminosity for the FERO sample, highlighting those sources for which a
significant relativistic broad Fe K$_{\rm \alpha}$ line detection has been
found. All broad Fe line detections correspond to objects with luminosities
below $\sim$ 10$^{44}$~erg s$^{-1}$. To investigate a possible dependence of
relativistic broadening on the source luminosity, the following three
luminosity bins are defined: L1: L$_{\rm X}$ $<$ 0.25 $\times$10$^{44}$~erg
s$^{-1}$; L2: 0.25 $\times$ 10$^{44}$ $<$ L$_{\rm X}$ $<$ 1.20
$\times$10$^{44}$~erg s$^{-1}$; L3: L$_{\rm X}$ $>$ 1.20 $\times$ 10$^{44}$
ergs$^{-1}$. Considering the FERO sample, each bin was chosen to contain the
same number of sources ($\sim$50), with no further distinctions (see
Table~\ref{tab_frac_det}). No significant difference is found between the
number of sources with a relativistic line when comparing the lowest
luminosity bin L1 to the intermediate luminosity bin L2, either using the
sources from the FERO (0.7$\sigma$) or flux-limited (0.5$\sigma$) sample. The
highest luminosity bin L3 contains no detections. However, with the
uncertainties in the detection fraction in the L3 luminosity bin (see
Table~\ref{tab_frac_det}), it is possible to estimate the difference in
detection fraction between the L1 and L3 luminosity bins, at the 2.2$\sigma$
and 0.9$\sigma$ levels for the FERO and flux-limited samples, respectively. To
remove the dependency with luminosity and check whether an intrinsic
difference exist between Seyfert galaxies, and quasars in terms of detection
fraction, a luminosity range containing the same number of Seyfert galaxies, and
quasars is selected (L: 0.15 $\times$ 10$^{44}$ $<$ L$_{\rm X}$ $<$
2.69$\times$10$^{44}$~erg s$^{-1}$, Table~\ref{tab_frac_det}). The detection
fractions are 10$\pm$7\% (4/40) and $\leq$10\% (2/40) for Seyfert galaxies and
quasars, respectively, for sources in the FERO sample, and 18$\pm$14\% (3/17)
and $\leq$92\% (1/3) for sources in the flux-limited sample. The low number of
detections prevents any statistically significant results from being
drawn. Table~\ref{tab_frac_det} summarizes the different detection fraction
for the different source and luminosity class within the whole FERO and
flux-limited samples. The correlation between the Fe K$_{\rm \alpha}$ line EW
and the hard X-ray luminosity based on individual source detections and within
the flux-limited sample is investigated in Section~\ref{LineParameters}. This
analysis is extended to the remainder of the FERO sample in the companion
paper by Longinotti et al.

\begin{table}
\caption{Fraction of sources in the FERO sample with a significant
relativistic broad Fe K$_{\rm \alpha}$ line EW detection for different source type
and luminosity class.}
\label{tab_frac_det}      
\centering          
\begin{tabular}{l c c c}
\hline\hline                             
\multicolumn{1}{c}{Obj. Class/} & \multicolumn{2}{c}{Sample} \\ 
\multicolumn{1}{c}{Luminosity}  & \multicolumn{1}{c}{FERO} &
\multicolumn{1}{c}{Flux Limited}\\
& \multicolumn{1}{c}{(\%)} & \multicolumn{1}{c}{(\%)} \\

\hline                    
Sy+QSO      &    9$\pm$3 (13/149)& 36$\pm$14 (11/31) \\
\hline                    
Sy          &   13$\pm$5 (11/82) & 36$\pm$15 (10/28)\\
QSO         &    $\leq$6 (2/67)  &  $\leq$92 (1/3)  \\
\hline
NL          &   13$\pm$9 (4/30)  & $\leq$100  (2/4) \\
BL          &   12$\pm$6 (7/60)  & 35$\pm$18 (7/20) \\
\hline    
L1 (Sy+QSO) &   16$\pm$7 (8/50)  & 44$\pm$22 (8/18) \\
L2 (Sy+QSO) &   10$\pm$6 (5/50)  & 27$\pm$23 (3/11) \\ 
L3 (Sy+QSO) &    $\leq$2 (0/49)  &  $\leq$47  (0/2) \\
\hline
L Sy        &    10$\pm$7 (4/40)  & 18$\pm$14 (3/17) \\
L QSO       &    $\leq$10 (2/40)  &  $\leq$92 (1/3) \\ 
\hline                  
\end{tabular}
\tablefoot{
Sy: Seyfert galaxy, QSO: quasar; NL: narrow line, BL:
broad line; L1: L$_{\rm X}$ $<$ 0.25 $\times$ 10$^{44}$~erg s$^{-1}$; L2: 0.25
$\times$ 10$^{44}$ $<$ L$_{\rm X}$ $<$ 1.20 $\times$ 10$^{44}$~erg s$^{-1}$; L3:
L$_{\rm X}$ $>$ 1.20 $\times$ 10$^{44}$~erg s$^{-1}$, where L1, L2 and L3 were
chosen to contain the same number of sources ($\sim$50); L: 0.15 $\times$
10$^{44}$ $<$ L$_{\rm X}$ $<$ 2.69 $\times$10$^{44}$~erg s$^{-1}$, L was
chosen to contain the same number of Seyfert galaxy and quasars (40) (see text for
details).
}
\end{table}

\subsection{Average properties of the relativistic broad Fe K$_{\rm \alpha}$ line within the flux-limited sample}\label{LineDetections}

\par In the next two sections, the average properties of the relativistic
broad Fe K$_{\rm \alpha}$ line are discussed, first, in terms of the line
equivalent width and second, in terms of the disc parameters derived from the
{\it Kyrline} model. To derive meaningful results, only the 31 sources in the
flux-limited sample are considered; therefore, information from the 11 sources
(the first 11 sources listed in Table~\ref{tab_detections}) with a broad Fe
K$_{\rm \alpha}$ line detection and the 20 sources with an upper limit of the
broad Fe K$_{\rm \alpha}$ line are used (see Table~\ref{FL_UL}).

\subsubsection{Line equivalent width}\label{LineDetectionsEW}

\par For each source with a significant relativistic broad line detection, the
measured value of the broad Fe K$_{\rm \alpha}$ line EW is always below 300~eV
(see Table~\ref{tab_detections}). Considering all the detected lines, the mean
value of the distribution of broad Fe K$_{\rm \alpha}$ line EW is
143$\pm$27~eV, with a 1$\sigma$ standard deviation of 91~eV. It has to be kept
in mind that the derived value for the mean is likely to overestimate the {\it
true} mean of the distribution. The explanation can be found in the fact that
the parent distribution of line EW is truncated by the ability to detect weak
lines, which can only be found in sources with very a high signal-to-noise
spectrum (see Figure~\ref{fig_detections}). 

\par To derive a more representative value of the mean EW, the weighted mean
of the Fe K$_{\rm \alpha}$ broad line EW distribution was calculated using the
1$\sigma$ statistical errors on the line EW for the individual weights. In
this way, a weighted mean value of 76$\pm$3~eV with 1$\sigma$ standard
deviation of 39~eV is obtained. This value is lower than the mean value
derived above, and consistent with it to within $\sim$1.3$\sigma$. There are
two important caveats regarding the weighted mean. One, the calculation of the
weighted mean assumes that the individual values of the line EW share the same
underlying distribution and that such distribution is Gaussian, which does not
necessarily have to be the case. Second, the use of the inverse square of the
individual errors in the calculation of the weighted mean implies that, for a
set of measurements with the same relative errors, the weighted mean is biased
towards the lower values in the set. With a few exceptions, the relative
errors in the line EW are comparable regardless of the value of the line EW,
which implies that lower values of the line EW contribute more to the weighted
mean. As a result, one could consider these two values, i.e. the mean and the
weighted mean, one biased towards higher EW values and the other towards lower
EW values, as the limits where the {\it true} mean of the distribution of
relativistic broad Fe K$_{\rm \alpha}$ line EW in AGN would be
expected. Table~\ref{tab_mean_ew} summarizes these results.

\begin{table}[t]
\caption{Weighted mean ($<$EW$>_w$) and 1$\sigma_w$ standard deviation, and
  mean ($<$EW$>$) and 1$\sigma$ standard deviation, of the Fe K$_{\rm \alpha}$
  line EW distribution.}
\label{tab_mean_ew}      
\centering          
\begin{tabular}{l c c c c}
\hline\hline                             
Source  & \multicolumn{1}{c}{$<$EW$>_w$} & \multicolumn{1}{c}{$\sigma_w$} &
  \multicolumn{1}{c}{$<$EW$>$} & \multicolumn{1}{c}{$\sigma$} \\ 
Class  & & & & \\
       & \multicolumn{1}{c}{(eV)}       & \multicolumn{1}{c}{(eV)}     & 
  \multicolumn{1}{c}{(eV)}     & \multicolumn{1}{c}{(eV)} \\
\hline
SY+QSO (11) &   76$\pm$3  & 39 & 143$\pm$27  & 91 \\
SY     (10) &   72$\pm$3  & 37 & 141$\pm$30  & 95 \\
QSO    (1)  &  170$\pm$22 & $----$ & 170$\pm$22  & $----$\\
\hline
\end{tabular}
\tablefoot{
Only sources with an Fe K$_{\rm \alpha}$ broad line detection
within the flux-limited sample were used to compute these values. A
distinction has been made according to source class. The number of sources
used in each case to derive these values is in parenthesis).
}
\end{table}

\begin{figure}
  \centering
  \includegraphics[width=.45\textwidth]{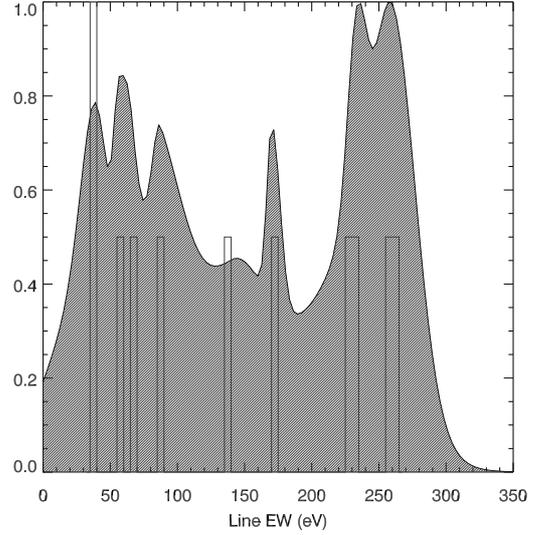}\\
  \caption{Normalised distribution (shaded area) for the line EW for sources
  with a significant broad Fe K$_{\rm \alpha}$ line detection within the
  flux-limited sample. The 1-dimensional histograms of this distribution is
  overlaid (white histogram).}
  \label{EW_Distributions}
\end{figure}

Figure \ref{EW_Distributions} shows the normalised distribution of the
line EW for sources with a significant broad Fe K$_{\rm \alpha}$ line
detection within the flux-limited sample. This distribution is generated by
combining one Gaussian distribution per measurement, where each Gaussian takes
the mean and standard deviation of the corresponding best-fit-parameter value
and its error at the 1$\sigma$ level (upper or lower limits are not present in
this plot). Overlaid on top of the Gaussian distribution is the
corresponding 1-dimensional histogram, highlighting the contribution of each
individual measurement to the overall distribution. Both distributions are
normalised to their respective peak values.

\subsubsection{Disc parameters}\label{LineDetectionsPar}

\par
\begin{figure*}
  \centering
  \includegraphics[width=.45\textwidth]{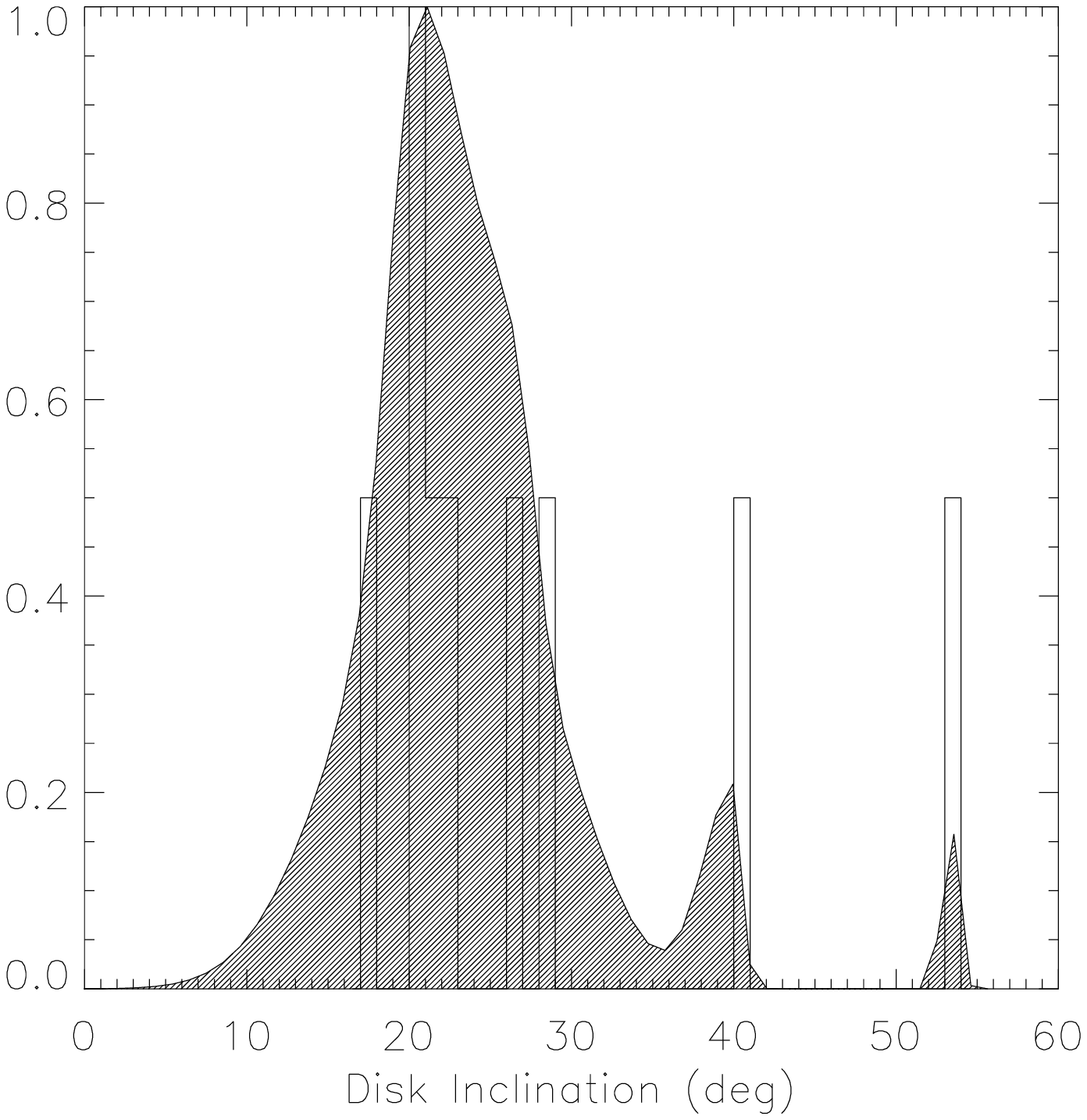}
  \includegraphics[width=.45\textwidth]{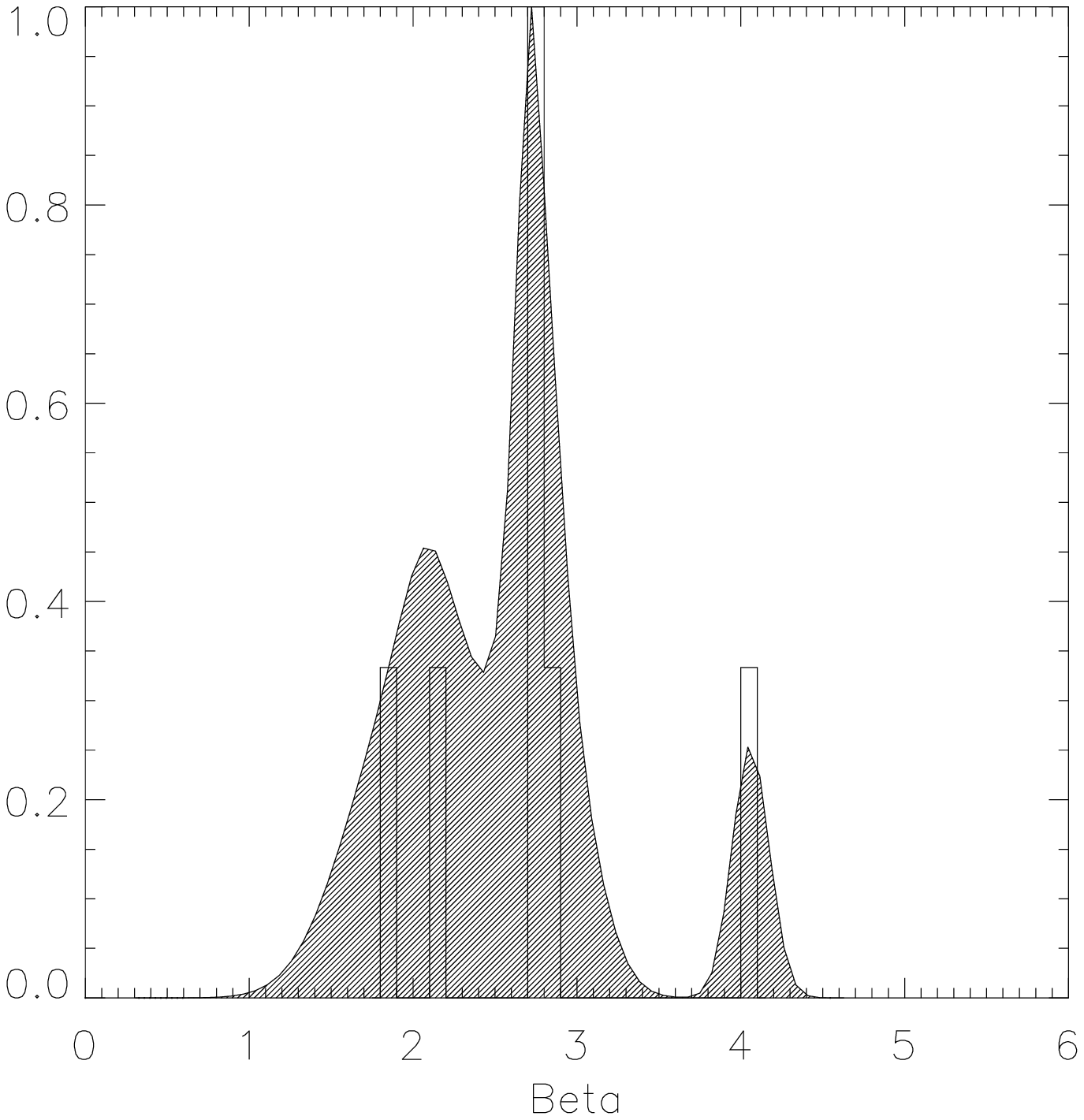}\\
  \caption{Normalised distributions (shaded area) for the disc inclination
  angle (left) and disc emissivity (right) for sources with
  a significant broad Fe K$_{\rm \alpha}$ line detection within the
  flux-limited sample. The 1-dimensional histograms of these distributions are
  overlaid (white histograms).}
  \label{Disc_Distributions}
\end{figure*}

 The {\it Kyrline} model parameters $\theta$, $\beta$, and {\it a}
respectively correspond to the disc inclination angle with respect to the
observer, the power-law index of the radial disc emissivity law, and the black
hole spin, and together they provide important information on the properties
and conditions of the accretion disc. These values are reported in
Table~\ref{tab_detections} for the sources with a significant broad Fe K$_{\rm
\alpha}$ line detection. For some sources, these parameters could not be
constrained in the spectral fit and in these cases lower or upper limits are
reported. Using only the parameters from the 11 line detections within the
flux-limited sample, the mean and 1$\sigma$ standard deviation values of
$\theta$, $\beta$ and {\it a} are derived via the censored mean method
described in Bianchi et al. (2009a). While these authors include measurements
and upper limits, lower limits are also considered
here. Table~\ref{model_mean} lists these mean values and the 1$\sigma$
standard deviation obtained using this method. (With only two spin
measurements, the mean black hole spin is omitted.)
Figure~\ref{Disc_Distributions} shows normalised distributions for $\theta$
and $\beta$. As for the case of the distribution of the line EW in
Section~\ref{LineDetectionsEW}, these distributions are generated by combining
one Gaussian distribution per measurement. The black hole spin is constrained
well only in two cases, MCG-6-30-15 and MRK509, with a best-fit value of 0.86
$(^{+0.01}_{-0.02})$ and 0.78 $(^{+0.03}_{-0.04})$, respectively.

\par The mean value of $\beta$ is about 2.4, but with a wide spread of
values. Among the 11 sources, 6 have values of $\beta$ higher than and
inconsistent with 2, which means that emission is dominated by the innermost
accretion disc regions. In three sources $\beta$ is definitely lower than 2,
and the emission is therefore dominated by the outermost regions. It must be
recalled that a power law is often a poor approximation of the real emissivity
law, which depends on the (unknown) geometry of the emitting corona. (See for
instance Martocchia et al. 2000 and Martocchia et al. 2002 for the
emissivity laws in the {\it lamp-post} geometry.) It is, however, worth noting
that in two sources, MCG-6-30-15 and MRK~509, $\beta$ is about 4, or larger,
which, under the {\it lamp-post} assumption, would imply an emitting source
located at only a few gravitational radii from the black hole
(\cite{Martocchia2002}). In addition and for the case of MCG-6-30-15,
\cite{Wilms2001} argue that this is not sufficient to explain the high $\beta$
value and conclude that it is necessary to invoke magnetic extraction of the
black hole spin energy to explain it (\cite{Blandford77}).

\par Not much can be said about the spin distribution. In five cases this
parameter is totally unconstrained, since any permitted value of {\it a} is
acceptable, and in another case it is rather poorly constrained ($a>0.16$). In
three cases, lower limits of about 0.5 can be placed, while in the remaining
two sources, the measurements imply high values of the spin, but inconsistent
with maximally rotating. It is interesting to note that, whenever a constraint
can be placed, it always implies the rejection of the static black hole
solution. However, this is a likely observational bias since non rotating
black holes produce narrower lines, which are easier to detect with respect to
the underlying continuum than lines produced by maximally spinning black
holes, where the red tail of these lines can be confused with the underlying
continuum. A maximally spinning black hole solution is statistically
favoured than a static black hole solution. Given how few sources there are,
the approximations made (see for instance the above discussions on $\beta$)
and the subtle interplay between the various disc parameters, it is however
premature to draw any firm conclusion in this respect.

\begin{table}
\caption{Mean and 1$\sigma$ standard deviation of the best-fit {\it Kyrline}
  model parameters for sources with a relativistic broad Fe K$_{\rm \alpha}$ line
  detection.}
\label{model_mean}      
\centering 
\begin{tabular}{l c c}     
\hline\hline                             
Line Parameter   & Mean    & $\sigma$ \\
({\it Kyrline})  &       &          \\ 
\hline                    
$\theta$($^\circ$) & 28$\pm$1    &     5 \\
$\beta$            & 2.4$\pm$0.1 &   0.4 \\
\hline
\end{tabular}
\tablefoot{
With only two spin measurements, the mean black hole spin is
omitted.}
\end{table}

\par
\begin{figure*}
  \centering  
  \includegraphics[width=.45\textwidth]{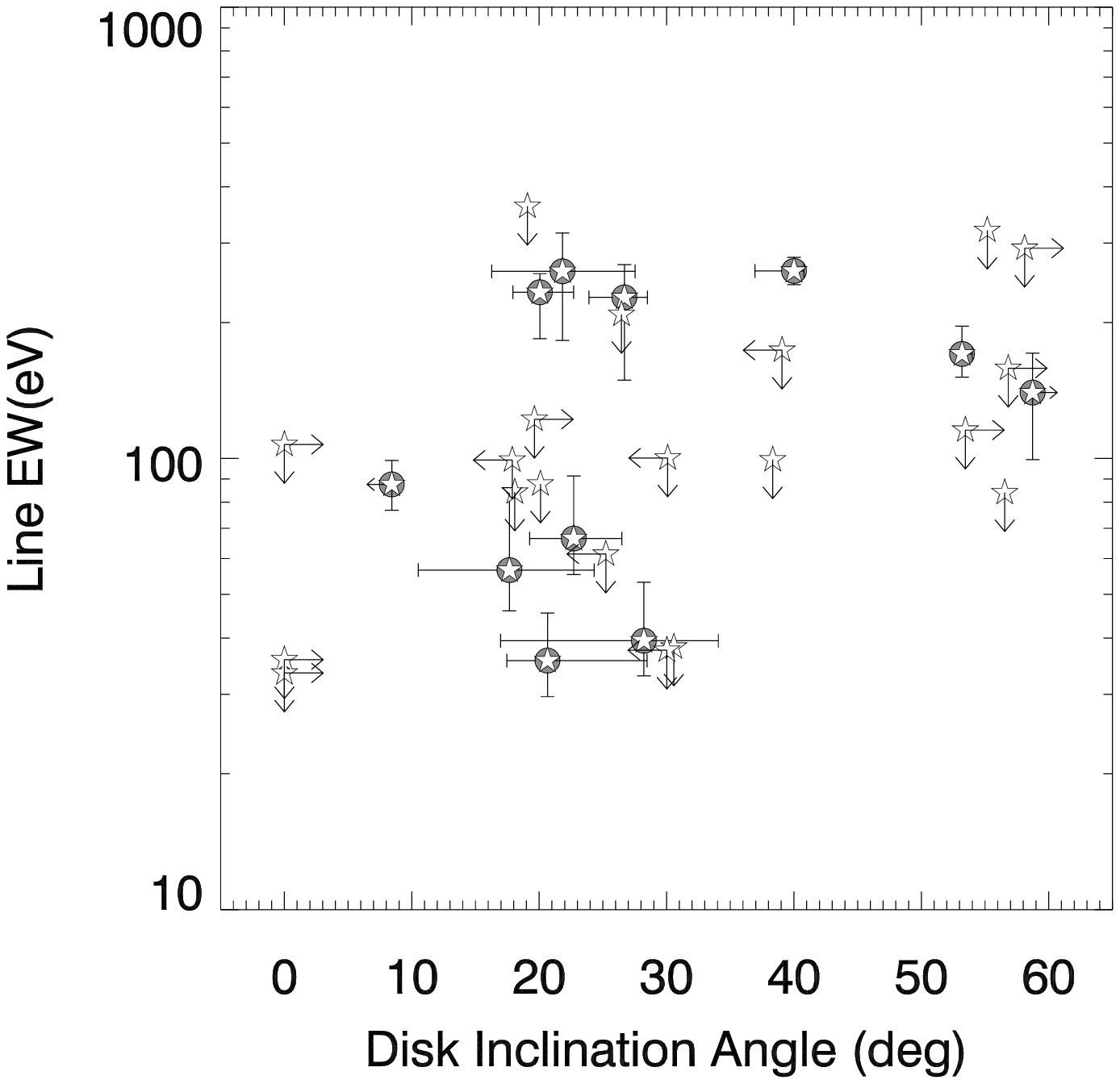}
  \includegraphics[width=.45\textwidth]{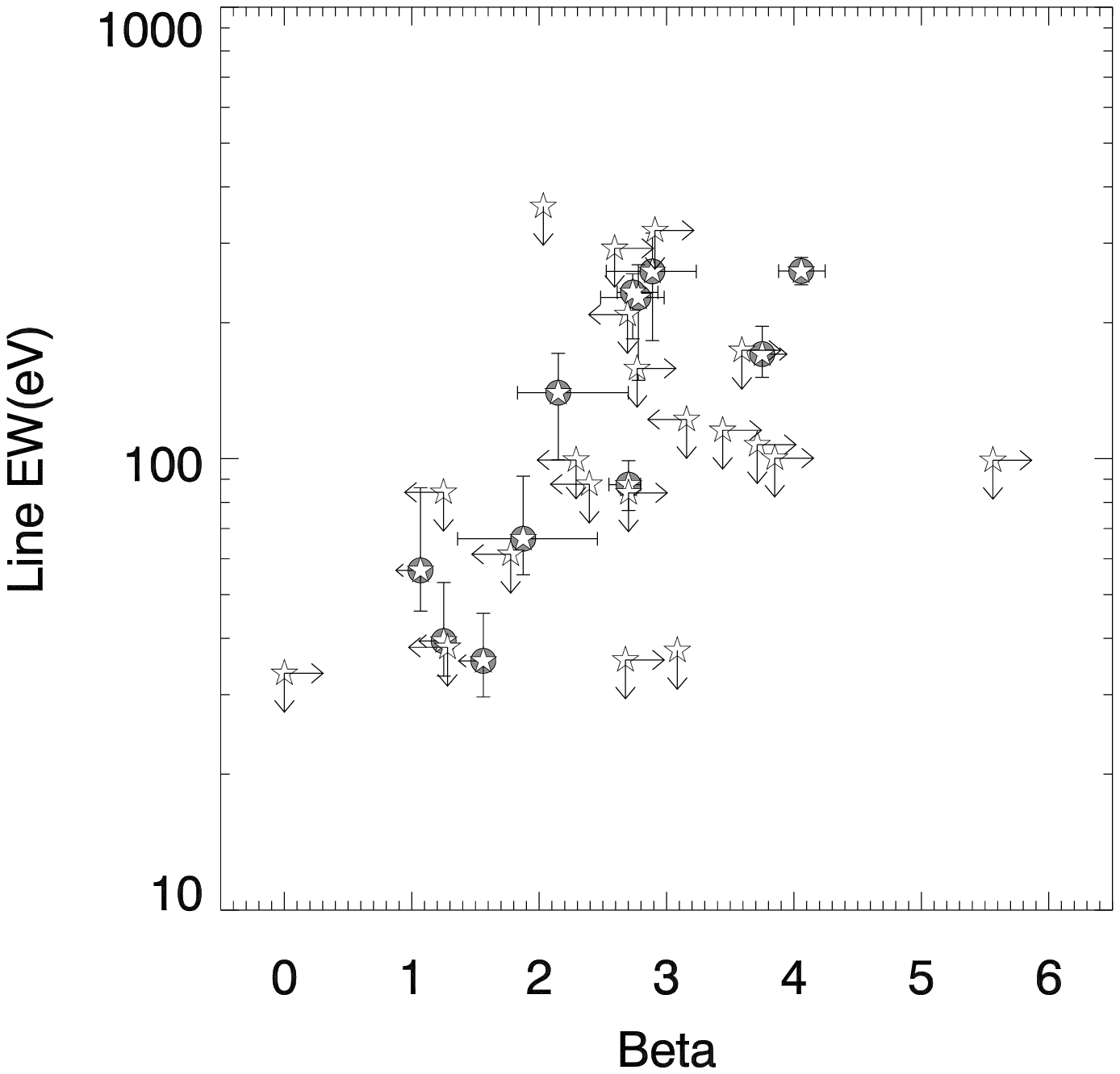}
  \caption{Dependence of disc inclination angle and emissivity with the
  relativistic broad Fe K$_{\rm \alpha}$ line EW. Filled circles indicate line
  detections at $\ge$5$\sigma$ confidence level (where error bars indicate the
  90\% confidence level intervals), line upper limits are given at the 90\%
  confidence level. {\it Kyrline} parameters upper and lower limits are given
  at the 90\% confidence level.}
  \label{EW_Relations1}
\end{figure*}

 The relation of the relativistic broad Fe K$_{\rm \alpha}$ line intensity
with $\theta$, and $\beta$ is also plotted in Figure~\ref{EW_Relations1}. As
for black hole spin, constraints are too few (two sources only) to allow for
discussion of any possible correlation of the line EW with {\it a}.  These two
distributions only show two things. First, the distribution of Fe K$_{\rm
\alpha}$ line intensity with $\theta$ is consistent with the fact that
relativistic lines arising in accretion discs seen at high inclination angles
rather than from face-on discs are more difficult to detect for a given value
of the line EW (as noted by Nandra et al. 2007, see also Matt et
al. 1992). Second, the apparent lack of detections of small EW lines with
steep emissivity profiles is most likely an observational bias, and any real
correlation between the two physical quantities can not be claimed.

\subsection{Correlation studies within the flux-limited sample}\label{LineParameters}
 
\par Given all the upper limits derived for the line parameters, the censored
fit approach was adopted. This procedure for linear fitting is applicable to
data containing both measurements and upper and/or lower limits, and it is
based on the method used by \cite{Schmitt1985} and \cite{Isobe1986}. Here it
was applied in the same way as in Bianchi et al. (2009b). For each correlation
analysis, the mean Spearman's rank correlation coefficient and null hypothesis
probability were calculated. Only correlations with a null hypothesis
probability lower than 10$^{-3}$ (99.9\% c.l.) will be considered
statistically significant.

\par The correlations investigated in this work involve, on one hand, the
broad Fe K$_{\rm \alpha}$ line EW and the disc emissivity ($\beta$), and on
the other, some source physical properties such as the black hole mass
(M$_{\rm BH}$), the Eddington luminosity (L$_{\rm Edd}$), and the accretion
rate (expressed in terms of $\frac{L_{\rm bol}}{L_{\rm Edd}}$). Within the
censored fit method, the errors used for the line EW are 1$\sigma$ errors,
while the upper limits are at the 90$\%$ c.l.. In those
correlations involving $\beta$, the censored fit approach described in Bianchi
et al. (2009b) was slightly modified to account for the lower limits to the
$\beta$ parameter. The lower limit to $\beta$ can be interpreted simply as
consistent with 6 at the 90$\%$ c.l.. This value of 6 is the maximum value
allowed for $\beta$ in our model (see Section~\ref{Test_Mod_Par}).

\par The values of the black hole mass were taken from the CAIXA catalogue
(\cite{Bianchi2009a}), and they are available for 24/31 of the objects in the
flux-limited sample. (The reader is deferred to this paper for the references
from which these values have been extracted.) Using the black hole mass
(M$_{\rm BH}$), the Eddington luminosity was derived in this work according to
the usual definition: $\displaystyle{L_{\rm Edd}= (4\pi G M m_p c)\;/
\;\sigma_t = 1.26 \times 10^{38} \; (M_{\rm BH}/M_{\bigodot}) \; erg \;
s^{-1}}$. The accretion rate is considered as the ratio of the bolometric
luminosity to the Eddington Luminosity: $\displaystyle{\frac{L_{\rm
bol}}{L_{\rm Edd}}}$, where the bolometric luminosity has been derived by
applying the following bolometric correction: $\displaystyle{L_{\rm bol} = k_b
\times L_{\rm x}}$, where k$_b$ is the correction factor dependent on the
2-10~keV X-ray luminosity ($L_{\rm} x$) following \cite{Marconi2004}. The hard
X-ray luminosities derived in this work for those sources with a broad Fe
K$_{\rm \alpha}$ line detection are reported in Table~\ref{tab_objectinfo},
and they are in good agreement with those derived in the CAIXA catalogue.

\begin{table*}
\caption{Source physical properties used in this work for those sources with a
  relativistic 6.4~keV Fe K$_{\rm \alpha}$ line detection within the flux-limited
  sample.}
\label{tab_objectinfo}      
\centering 
\begin{tabular}{l c c | c c c}     
\hline\hline                             
\multicolumn{1}{c}{Source} & \multicolumn{1}{c}{L$_{\rm X}^{2-10~keV}$} &
\multicolumn{1}{c|}{Log$_{10}$(BH)} & \multicolumn{1}{c}{L$_{\rm bol}$} &
\multicolumn{1}{c}{L$_{\rm Eddington}$} & L$_{\rm bol}$/L$_{\rm Eddington}$  \\
\multicolumn{1}{c}{Name}   &  & & & & \\
       & \multicolumn{1}{c}{(10$^{44}$~erg s$^{-1}$)} & \multicolumn{1}{c|}{(M$_{\sun}$)} & \multicolumn{1}{c}{(10$^{44}$~erg s$^{-1}$)} & \multicolumn{1}{c}{(10$^{44}$~erg
s$^{-1}$)} &  \\ 
\hline                  
         IC4329A & 0.56 & 8.08  &15.97 &  7.42&  0.46 \\ 
     MCG-5-23-16 & 0.14 & 7.85  & 2.72 & 89.20&  0.03 \\ 
     ESO511-G030 & 0.22 & $---$ & 4.87 &$---$ &  $---$\\ 
     MCG-6-30-15 & 0.05 & 6.19  & 0.82 &  1.95&  0.42 \\ 
         NGC4051 & 0.003& 6.13  & 0.02 &  1.70&  0.01 \\ 
         NGC3516 & 0.03 & 7.36  & 0.49 & 28.86&  0.02 \\ 
         NGC3783 & 0.11 & 6.94  & 2.05 & 10.97&  0.19 \\ 
         NGC3227 & 0.01 & $---$ & 0.12 &$---$ &  $---$\\ 
          MRK509 & 1.04 & 7.86  &35.37 & 91.28&  0.39 \\ 
          MRK766 & 0.06 & 6.28  & 0.93 & 2.40 &  0.39 \\   
          ARK120 & 0.90 & 8.27  &29.58 & 234.6&  0.13 \\   	 
\hline
\end{tabular}
\end{table*}

\begin{figure*}
  \centering
  \includegraphics[width=.4\textwidth]{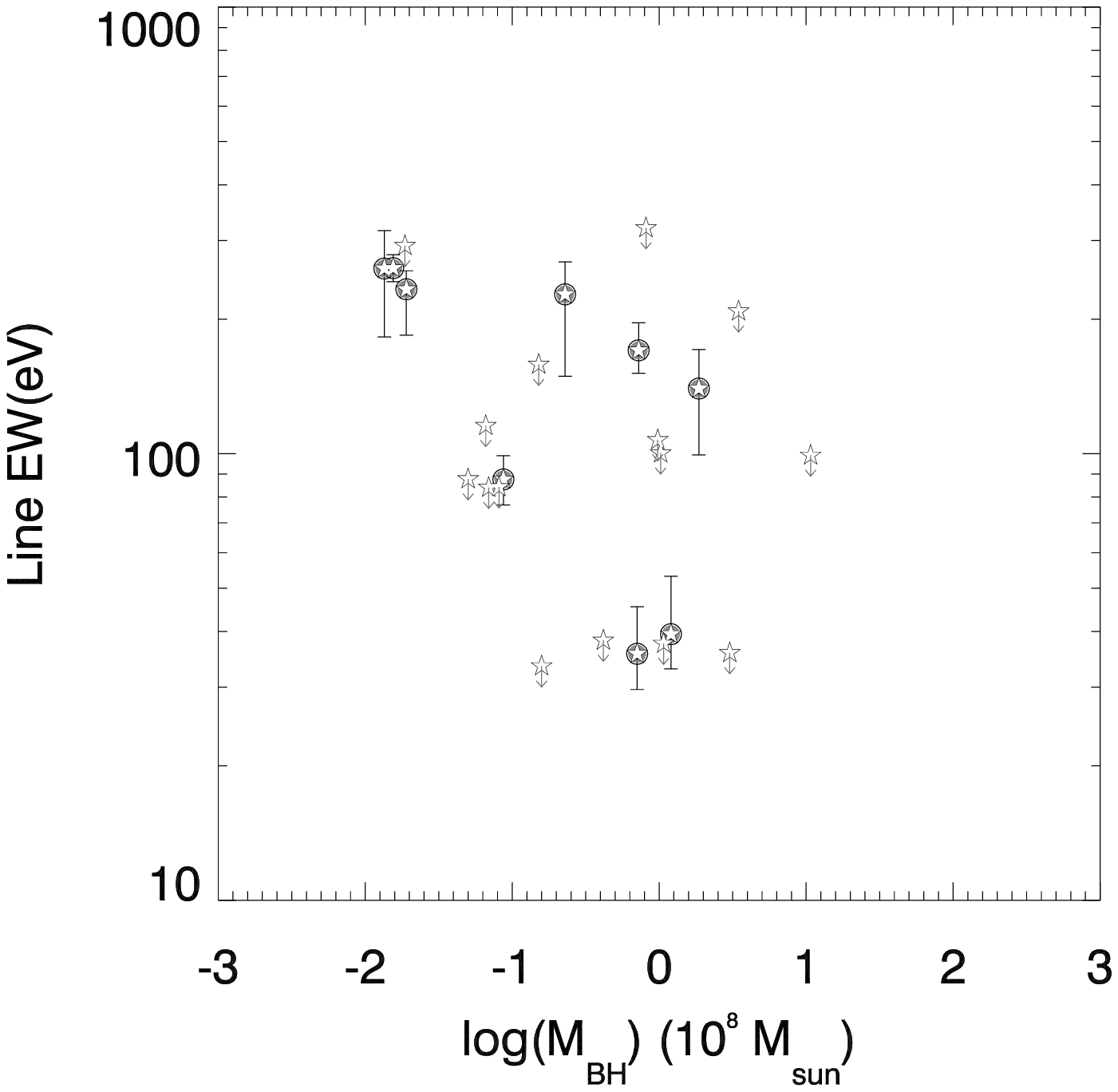}
  \includegraphics[width=.4\textwidth]{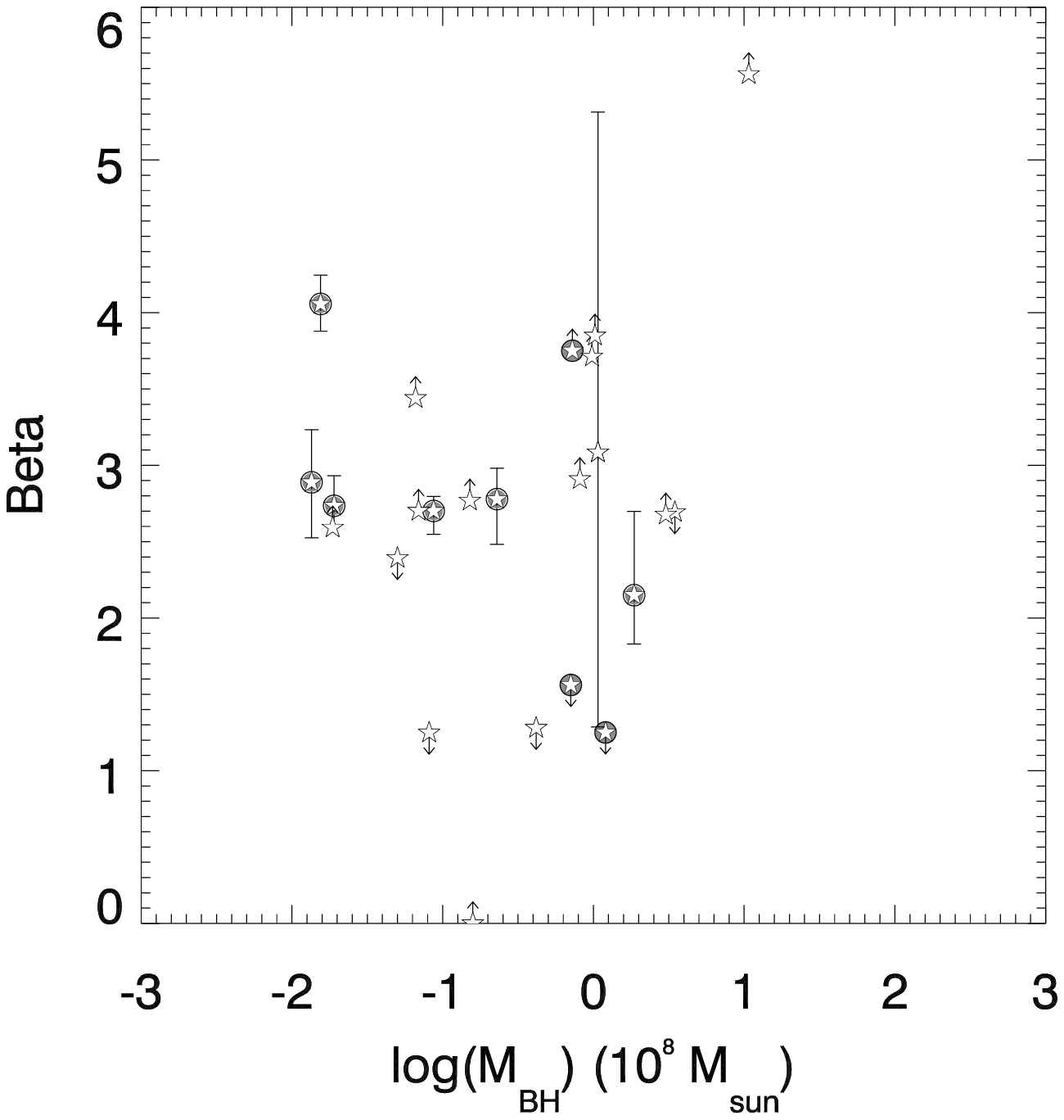}\\
  \includegraphics[width=.4\textwidth]{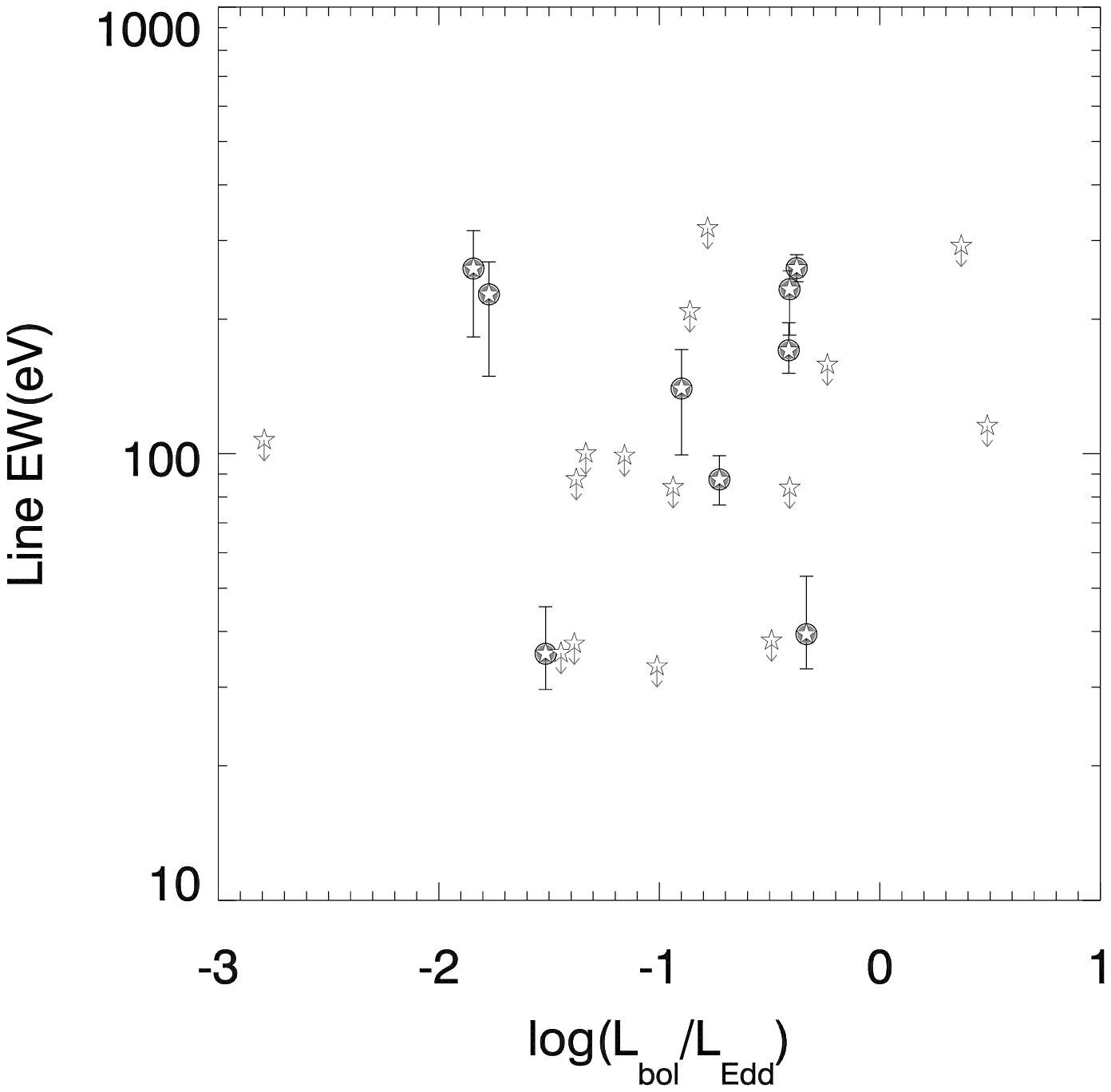}
  \includegraphics[width=.4\textwidth]{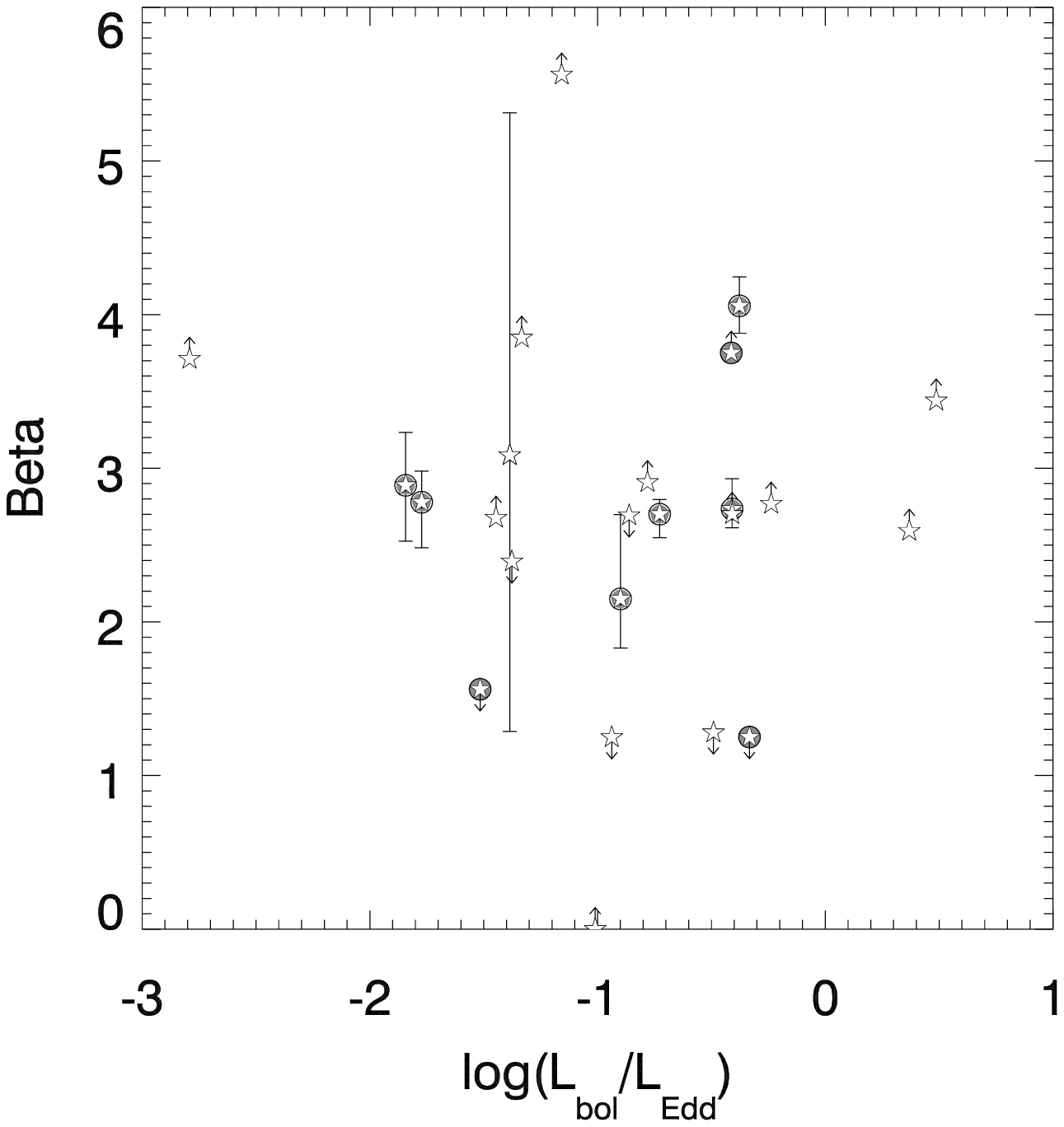}\\
  \includegraphics[width=.4\textwidth]{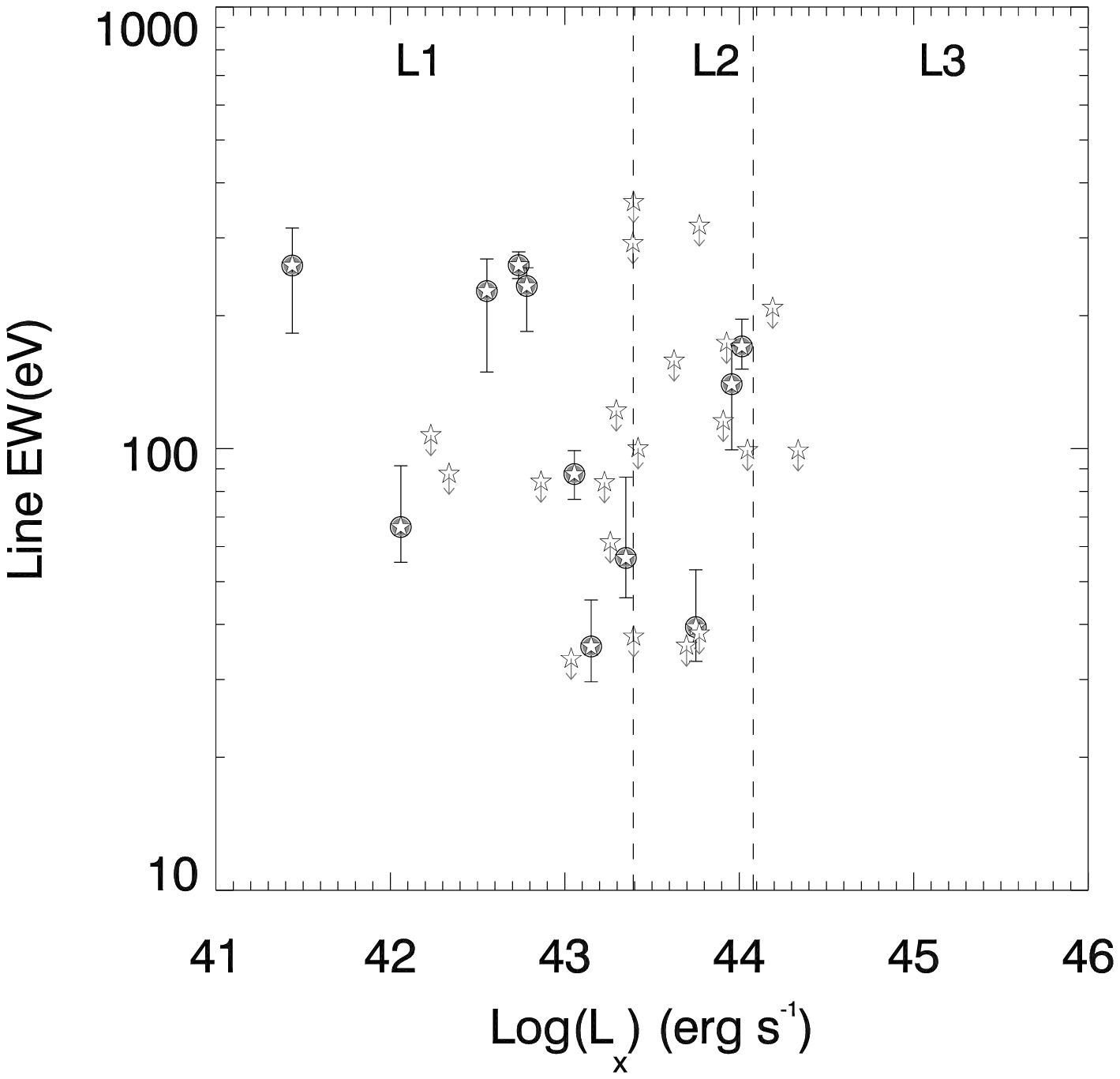}
  \includegraphics[width=.4\textwidth]{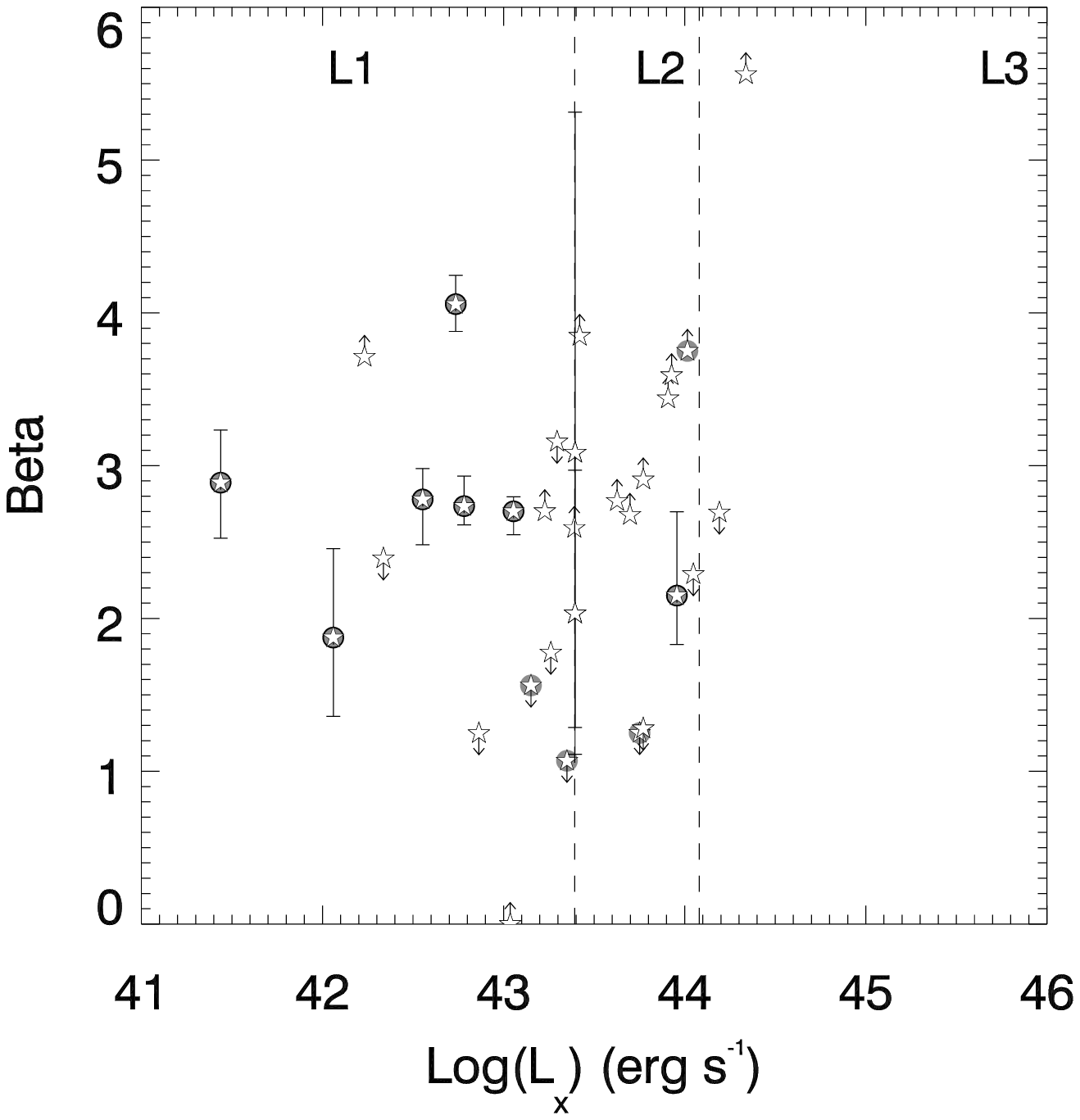}\\
  \caption{From top to bottom, relativistic broad Fe K$_{\rm \alpha}$ line
  equivalent width and $\beta$ vs. black hole mass, accretion rate and hard
  X-ray luminosity. Filled circles indicate broad Fe K$_{\rm \alpha}$ line
  detections at $\ge$5$\sigma$ confidence level (where error bars indicate the
  90\% confidence level intervals). The dashed vertical lines in the bottom
  plots indicate the L1, L2, and L3 luminosity bins as described in the text
  (see Section~\ref{fraction_det2}).}
  \label{EW_Relations2}
\end{figure*}

\par Table~\ref{rho_correlations} and Figure~\ref{EW_Relations2} summarize the
results of the correlation studies carried out. Table~\ref{rho_correlations}
lists the mean value of the Spearman's rank coefficient for the correlation of
the broad Fe K$_{\rm \alpha}$ line EW and $\beta$ with the three physical
parameters of interest, together with the corresponding null hypothesis
probability. A negative value of Spearman rho indicates anti-correlation
between the two parameters of interest. On the basis of these results, it is
concluded that no statistically significant correlation is found with any of
the considered physical properties of the AGNs.

\par Last, the relativistic Fe K$_{\rm \alpha}$ line EW derived in this work
and all the different parameters included in the CAIXA catalogue were
correlated (see Table 1 in \cite{Bianchi2009a}). No statistically significant
correlation has been found with any of the parameters, including the FWHM of
the H$_{\rm \beta}$ line and the narrow Fe 6.4~keV K$_{\rm \alpha}$ line
EW. However, the low statistics prevents us from drawing any robust physical
conclusions from the lack of such correlations.

\begin{table*}
\caption{Results of the correlation studies within the flux-limited
  sample.}
\label{rho_correlations}      
\centering 
\begin{tabular}{l c c c c}     
\hline\hline                             
Broad Fe K$_{\rm \alpha}$ Line EW (Y) vs. & a & b & Mean Spearman rho &  Null Hypothesis Probability \\ 
\hline                    
Black Hole Mass (X) & 1.64$\pm$0.12 & -0.20$\pm$0.11 & -0.34 & 0.14\\
Accretion Rate (X)  & 1.82$\pm$0.15 &  0.08$\pm$0.13 &  0.15 & 0.50 \\
X-ray Luminosity (X)& 7.62$\pm$5.16 & -0.14$\pm$0.12 & -0.13 & 0.51 \\
\hline\hline
Beta (Y) vs. & & & & \\
\hline
Black Hole Mass (X) &  0.36$\pm$0.11 & -0.02$\pm$0.09 & 0.05 & 0.75\\
Accretion Rate (X)  &  0.38$\pm$0.12 &  0.01$\pm$0.10 & 0.12 & 0.59\\
X-ray Luminosity (X)& -0.32$\pm$4.19 &  0.01$\pm$0.10 & 0.17 & 0.39\\
\hline
\end{tabular}
\tablefoot{Mean intercept (a) and slope (b) for the linear fits of a
$\log~(Y) = a + b \times \log~(X)$ function described in the text. Errors in
these two quantities are the statistical one standard deviation.}
\end{table*}

\section{Discussion}\label{Discussion}

\par Despite the fact that the first relativistically distorted line in an AGN
was observed about 15 years ago by ASCA (\cite{Tanaka1995},
\cite{Nandra1997}), and despite strong efforts to study them with more recent
observatories, like {\it XMM-Newton} and {\it Suzaku}, there are still large
uncertainties on the relative number of AGNs that possess such lines. The FERO
project has been designed to address the fundamental question of how often
these lines appear in AGN by estimating the fraction of objects
where relativistic effects are detected on a well-defined sample. This number
is crucial to establish if and how the standard AGN paradigm needs to be
modified.

\par A collection of 149 radio-quiet Type 1 AGN, all targeted by {\it
XMM-Newton}, has been assembled (the largest ever assembled for this kind of
study) in order to tackle the question posed above. One fundamental
requirement of the FERO project was to identify a complete and unbiased
subsample of sources with high signal-to-noise spectra to be able to derive
meaningful constraints on the physical properties of the sources. Only then it
would be possible to extrapolate the results derived from this subsample to
the unknown parent population of local radio-quiet AGN. A flux-limited sample
of 31 objects was defined by selecting the XSS sources (see
Section~\ref{FLSample}) with count rates in the 3-8~keV energy band greater
than 1 cts/sec that fulfill the FERO selection criteria (i.e. local Type 1
radio-quiet X-ray unobscured sources). In this section the statistical
properties of the flux-limited sample are discussed.

\par The baseline model used in FERO aims at providing the most accurate
representation of the physical ingredients that are typically found in an
average Type 1 AGN, so this model includes a Compton reflection component and
absorption by line-of-sight ionised gas for modelling the continuum, a series
of narrow Fe lines (5), and a relativistic Fe K$_{\rm \alpha}$ line component
fitted by the {\it Kyrline} model. The spin of the black hole, the radial
emissivity profile of the accretion disc, and the disc inclination angle to
the observer are free parameters in the fit of the line profile. To achieve a
uniform analysis over the whole FERO sample, no attempt has been made to find
the best-fit model on a source-by-source basis, nor has any attempt been made
to find a best-fit model by using the FERO sample as a whole. To reduce the
parameter space during the fitting procedure, several model parameters were
investigated to see whether they could be constrained due either to physical,
geometrical, or observational considerations. Tests were also carried out on
including different model components in the baseline model, and the results
have served either to consider them in the final baseline model or rule them
out. The most important of these tests was the replacement in the baseline
model of the neutral reflection component by an ionised reflection component,
with the consequent change in the relativistic line from neutral iron to ionised
iron. In fact, the results from these two scenarios were merged and presented as
if coming from a single one.

\par Last, a consistency test was carried out by using the hard
X-ray spectra of all the sources in the flux-limited sample, for which a
significant detection of the relativistic line can be claimed (those listed in
Table~\ref{tab_detections}), by describing it with the most self-consistent
reflection model envisaged, however keeping the model as simple as
possible. The comparison of the results of this more complex and
self-consistent model for the reflection components indicates that it does not
significantly affect the results derived from the more phenomenological
approach chosen for the baseline model adopted throughout the paper (refer to
Appendix~\ref{appendixf} for more details).

\subsection{Fraction of relativistic lines}\label{mainresults_frac}

\par The main result of the FERO analysis is the relativistic broad line
detection fraction, $36\pm14$\% in the flux-limited FERO subsample. The
observed detection fraction is regarded as a lower limit to the intrinsic
number of AGN that would show a broad Fe K$_{\rm \alpha}$ line if, for example, all
sources were observed with the same, and sufficient, signal-to-noise ratio. It
is important to stress once more that the observed detection fraction comes as
a result of adopting a rather conservative detection threshold of 5$\sigma$
confidence level, which is necessary since FERO does not focus on reporting
individual broad line detections in particular objects. Instead, it focuses
on statistical studies of the sample as a whole, which includes spectra of
very different statistical quality. Nevertheless, as a reference, by relaxing
this requirement to a 3$\sigma$ confidence level, the above detection fraction
rises to $61\pm20$\% (19/31).

\par Upper limits on the line EW can also provide important information. For
example, the fraction of sources in the flux-limited sample in which the broad
Fe K$_{\rm \alpha}$ line EW is limited below 100~eV is of the order of 50\%
(where 100\% of line EW UL are below 360~eV). This result points to a much
lower value of the average line intensity derived for local Type 1 AGN
compared to the one derived from the mean spectra of high redshift (z
$\gtrsim$ 1) Type 1 AGN ($<$EW$>$ $\sim$ 560~eV; \cite{Streblyanska2005}),
although there should be some caution in adopting this result since the
high mean EW reported could come as a result of the method adopted for co-adding
the spectra (\cite{Yaqoob2006}). The result derived in this work is also
consistent with the upper limit of $<$400~eV derived by Corral et
al. (2008). Using results on individual sources, the FERO analysis has
provided upper limits below 40~eV in four sources (MRK~279, NGC~5548,
ESO~198-G24, and MRK~590), which are constraining enough to have to stretch
disc-accretion theories by requiring the use of non-standard system
properties, such as very high disc inclination angles or element abundances
significantly lower than solar (\cite{George1991}).

\par The {\it non-detection fraction} of broad lines in the flux-limited
sample is 13\% (4/31). Inverting this argument, the presence of a broad K$_{\rm
\alpha}$ line cannot be excluded {\it a priori} in the remainder 87\%
(although the line is formally detected in 36\% (11/31) of the sources). The
value of 87\% can be regarded as the upper limit to the intrinsic fraction of
AGN where a relativistic Fe K$_{\rm \alpha}$ line with EW$>$40~eV could be
present in the parent sample. Nandra et al. (2007) have fully explored the range of
effects (geometry, ionisation, and element abundances) that would explain
why a broad line is seen in some objects but not in others, so they are not
repeated here.

\par A direct comparison of the results extracted from FERO with previous
works reveals that the detection fraction of 36\% derived for the flux-limited
sample is consistent with previous results, such as those reported in Nandra
et al. (2007) where a detection fraction of 35-40\% was obtained. When
compared to similar studies on sizable {\it XMM-Newton} samples
(e.g. \cite{Nandra2007}, \cite{Brenneman2009}), our sample differs in that our
sample is very well-defined and homogeneously selected. Moderately X-ray
obscured sources and radio-loud sources are both excluded from the FERO
sample. However, the most significant differences are the model used to fit
the relativistic profile, the fitting procedure, and the treatment of the
sources with multiple observations. For instance, Brenneman \& Reynolds (2009)
fitted the continuum model on the entire {\it XMM-Newton} energy band, but
they selected only one particular observation for each source, whereas Nandra
et al. (2007) include multiple observations of the same target without
combining them.

\subsection{Disc parameters}\label{mainresults_disc}

\par The second goal of the FERO project was to derive average line and system
properties and look into the question of whether a relativistic broad line can
be associated to any source physical properties and, if so, investigate this
dependence in more detail. The average broad line EW is of the order of
100~eV, however, a wide range of broad line EW (40-260~eV) are observed. As a
reference, a neutral reflector covering half of the sky, as seen from the X-ray
source ($\Omega /2\pi \sim 1$) and irradiated by a power-law spectrum with
$\Gamma \sim 1.9$, should produce an Fe K$_{\rm \alpha}$ line with equivalent
width from around 100 to 150~eV in the inclination range 0$^\circ$-60$^\circ$
and for solar abundances (\cite{George1991}). Therefore, line EWs measured
outside that range are likely to be produced mainly because i) the Fe
abundance is different from solar, and/or ii) the reflector (disc) subtends a
non-standard (smaller) solid angle as seen from the X-ray source. One physical
scenario where broad line EWs $\gtrsim$ 150~eV are plausible, involves
reprocess emission from a photoionised accretion disc (\cite{Matt1993}) or
under a geometrical scenario which involves anisotropic emission.

\par The shape of the broad lines observed in the X-ray spectra of AGN are a
potential carrier of information concerning the black hole spin. The issue of
placing constraints to the black hole spin has been pursued by several authors
(e.g. \cite{Nandra2007}, \cite{Brenneman2009}). Using different approaches,
both studies reached similar conclusions. While some objects show evidence of
a spinning black hole, with the present data it is very difficult to reject
the static solution. In this work the approach is that of modelling the
line with a code capable of including the spin as a free parameter. The {\it
Kyrline} model explicitly depends on the black hole spin. {\it Kyrline} is
part of a suit of routines called KY (\cite{Dovciak2004}), a code that
properly takes relativistic boosting, light bending, and time delays into account,
assuming an axysymetric accretion disc geometry surrounding a black hole. As a
general statement, the black hole spin is poorly constrained by the FERO
analysis, except for MCG-6-30-15 (0.86 $^{+0.01}_{-0.02}$) and MRK~509 (0.78
$^{+0.03}_{-0.04}$) where the {\it Kerr} black hole solution is favoured. In
five other sources, the static solution cannot be rejected, while in a
further five, the constraint implies values of the black hole spin above
$\sim$0.5.
 
\par Another aspect of the present study is to try to establish the physical
drivers behind the relativistic broad line by looking into any correlation
between the presence and/or properties of the broad line EW and sources
physical properties, such as black hole mass, X-ray luminosity, or Eddington
ratio. However, using only sources from the flux-limited sample, no
significant correlation has been observed between the broad Fe K$_{\rm
\alpha}$ line EW and these source properties. As previous works on large
samples have done (e.g. \cite{Iwasawa1993}, \cite{Bianchi2007},
\cite{Corral2008}), the FERO analysis has looked for inverse correlation
between Fe K$_{\rm \alpha}$ line emission and X-ray luminosity, the so-called
IT effect (previously known a X-ray Baldwin effect), where the EW of the
narrow Fe K$_{\rm \alpha}$ emission line is inversely proportional to the
2-10~keV X-ray luminosity. Earlier results on a smaller sample ($\sim$100 AGN)
using the spectra stacked residuals technique suggested that relativistic
broad profiles follow a similar trend, where broad profiles are more common in
low-luminosity AGN (\cite{Guainazzi2006}). With information from individual
source line profiles, the FERO analysis is unable to confirm such results,
finding no statistically significant indication for the IT effect on broad Fe
lines. However, it is worth pointing out that the lack of sources in the
flux-limited sample with luminosities above 10$^{44}$~erg s$^{-1}$ prevents
any firm conclusions. The IT effect will be investigated using the spectra
stacked residuals technique using those sources in the FERO sample with an
upper limit to the relativistic Fe K$_{\rm \alpha}$ line EW (Longinotti et
al. {\it in preparation}).

\subsection{FERO in a wider context}\label{mainresults_fero}

\par An important remark should be made about the relativistic line in sources
that are characterized by other phenomena. A recent paper by
\cite{DeMarco2009} carried out a systematic search for the
presence of red- or blue-shifted components in the time-averaged spectra of the
sources in the flux-limited sample. The study also looks into the spectral
variability of these spectral components. A direct comparison with their
results shows that 10 out of the 11 sources ($\sim$90\%) within the
flux-limited sample (all except ESO 511-G030) with a relativistic broad line 
indicate either red- or blue-shifted components, or both. In 6 out of
these 10 sources (IC4329A, NGC~3783, NGC~3516, MRK~509, MCG-6-30-15, and
MRK~766) there is evidence of significant variability ($>$90\%) in the excess
map (5.4-7.2~keV).

\par From the literature on the warm absorber phenomenon, we note that 9 out
of 11 sources with line detections have warm absorbers: NGC3783, NGC3516,
MRK~509, MCG-6-30-15, MRK~766, NGC~4051 (\cite{Blustin2005}; McKernan et
al. 2007), IC4329A (\cite{Steenbrugge2005}), NGC~3227 (\cite{Markowitz2009}).
In the case of ESO~511-G030 the presence of line-of-sight ionised
absorption has not been ascertained yet (Longinotti private communication).
The only 2 sources in the flux-limited sample with no warm absorber and no
transient features are Ark~120 and MCG-5-23-16, so it is tempting to deem
these two sources as the most robust detections, due to the lack of an
alternative scenario to the relativistic line.

\par That sources with significant relativistic lines simultaneously show Fe K
band variability and a warm absorber is probably twofold. There might be a
physical connection between, e.g., the presence of the relativistic line and
of the transient features, with some variability associated to it.  But, it
could well be the result of an observational bias, because these sources were
extensively (and repeatedly) observed because they were known to be
interesting {\it a priori}, so they ended up with the best signal-to-noise
data and this explains the coincidence of the broad Fe line, warm absorber,
and transient features in the same sources.

\par To conclude, it goes without saying that the FERO project would benefit
from an expansion in several directions. The most obvious would be to complete
the observations of the flux-limited sources such that the number of X-ray
counts in the 2-10~keV fall above 1.5$\times$10$^5$~cts. The expansion of the
number of sources with 2-10~keV X-ray luminosities above 10$^{44}$~erg
s$^{-1}$ would also allow studies like the IT effect to be carried out at a
significant level and, of course, to investigate other correlations where the
X-ray luminosity plays an important role.

\par Another crucial aspect to consider in future studies
regarding relativistic disc lines is broadband coverage. This would allow
exploration of the spectral regime above 10~keV and the opportunity to
discern whether the spectral contribution at these energies can be attributed to a
reflection component off distant matter and/or the accretion disc, analogous
to the case of MCG-5-23-16 (\cite{Reeves2007}). The importance of broadband
coverage is also highlighted by, for example, the recent detection of a broad
Fe K$_{\rm \alpha}$ line in Fairall 9 in Suzaku observations, which have allowed a
tight constraint on the black hole spin (\cite{Schmoll2009}).

\par The FERO project has highlighted the importance of the availability of a
complete and {\it well-exposed} AGN sample to shed some light on the origin
and properties of the relativistic broad Fe lines in AGN, the environment, and
conditions in which they are produced, and to provide better understanding
of the scientific drivers needed for future high-throughput X-ray missions,
such as IXO.

\section{Conclusions}\label{Conclusions}

\par The main conclusions that are drawn from the results of the systematic
and homogeneous spectral analysis of the large number of sources in FERO can
be summarised as follows.

\begin{itemize}

\item[$\bullet$]{The observed fraction of sources in the FERO sample that
present strong evidence ($\ge$5$\sigma$ significance) of a relativistic broad
Fe K$_{\rm \alpha}$ line is of the order of 9$\%$ (13/149). Considering only
sources from the flux-limited sample, the detection fraction rises to around
36$\%$ (11/31). This number can be interpreted as a lower limit to the
intrinsic fraction of AGN with relativistic Fe line broadening. The sources
with a significant Fe K$_{\rm \alpha}$ line detection mostly belong to sources
with good statistical quality spectra ($\gtrsim$1.5$\times$10$^5$~cts), so it
is not surprising that the majority belong to the flux-limited
sample. Considering only sources from the flux-limited sample, a broad Fe
K$_{\rm \alpha}$ line at the level of 40~eV can be rejected in 4 objects,
87$\%$, which can be regarded as an upper limit to the intrinsic fraction of
AGN with relativistic Fe K$_{\rm \alpha}$ line broadening.}

\item[$\bullet$]{There is no significant difference between Seyferts and
quasars or narrow line and broad line objects in terms of detection fraction.}

\item[$\bullet$]{All sources with a broad Fe K$_{\rm \alpha}$ line detection
have a hard X-ray luminosity in the 2-10 keV energy band below $\sim$1
$\times$ 10$^{44}$ erg s$^{-1}$. The difference in the detection fraction
between the highest and lowest luminosity bins as defined in this work is only
2$\sigma$.}

\item[$\bullet$]{From those sources with a significant Fe K$_{\rm \alpha}$ line
  detection within the flux-limited sample, it is found that:}
\begin{itemize}
\item[$\circ$]{The average relativistic Fe K$_{\rm \alpha}$ line EW is of
  the order of 100~eV (never higher than 300~eV for any given source).}
\item[$\circ$]{The average system properties and 1$\sigma$ standard deviation 
  inferred from the {\it Kyrline} model can be summarised as:\\
  $<\theta>$  = 28 $\pm$ 5$^\circ$, consistent with an intrinsic random distribution
  of inclination angles.\\
  $<\beta>$  = 2.4 $\pm$ 0.4, with a wide spread of values.\\
  The spin value {\it a} is poorly constrained, except for MCG-6-30-15 (0.86 $(^{+0.01}_{-0.02})$)
  and MRK509 (0.78 $(^{+0.03}_{-0.04})$) where the {\it Kerr} black hole solution is favoured.}
\item[$\circ$]{No significant trend is found between the Fe K$_{\rm \alpha}$
  line EW and any of the line parameters investigated ($\theta$, $\beta$, and
  {\it a}).}
\item[$\circ$]{No significant correlation has been found between the Fe
  K$_{\rm \alpha}$ line EW or disc emissivity ($\beta$) and the source physical
  properties investigated, such as, black hole mass, accretion rate, and hard
  X-ray luminosity.}
\end{itemize}
\end{itemize}

\begin{acknowledgements}
This work is based on observations by {\it XMM-Newton}, an ESA science mission
with instruments and contributions directly funded by ESA member states and
NASA. I. de la Calle would like to acknowledge support by the Torres Quevedo
fellowship from the Ministerio de Ciencia e Innovaci\'on Espa\~nol and INSA.
A.L. Longinotti acknowledges travel support provided by NASA through the
Smithsonian Astrophysical Observatory (SAO) contract SV3-73016 to MIT for
Support of the Chandra X-Ray Center, operated by SAO for and on behalf of NASA
under contract NAS8-03060. We acknowledge financial support from the Faculty
of the European space Astronomy Centre (ESAC). G. Miniutti thanks the Spanish
Ministerio de Ciencia e Innovaci\'on Espa\~nol and CSIC for support through a
Ram\'on y Cajal contract. P.O. Petrucci acknowledges support by the PCHE
(Ph\'enom\`enes Cosmiques des Hauutes Energies) French working
group. S. Bianchi, G. Matt and E. Piconcelli acknowledge financial support
from Italian Space Agency, under grant ASI I/088/06/0. G. Ponti thanks ANR for
support (ANR-06-JCJC-0047). We thank the referee for a careful reading of the
manuscript and the useful comments that helped to improve it.
\end{acknowledgements}

\begin{appendix}

\newpage
\section{Sources from the flux-limited sample.}\label{appendixa}

\par List of 33 sources in the flux-limited sample. These sources have been
selected from the RXTE all-sky Slew Survey (XSS, \cite{Revnivtsev2004}) with a
count rate in the 3-8~keV energy band greater than 1~cts/sec and fulfilling
the FERO source selection criteria. The XSS is nearly 80\% complete at the
selected flux level for sources with Galactic latitude greater than
10$^\circ$. For two of them, UGC~10683 and ESO~0141-G055, no {\it XMM-Newton}
data were available as of April 2008. This leaves the number of XSS-selected
bright sources in the FERO sample at 31.

\begin{table*}
\caption{Flux-limited sample: 33 sources from the RXTE Slew Survey with a
count rate in the 3-8~keV energy range greater than 1 cts/sec and fulfilling
the FERO source selection criteria.}
\label{tab_fls_src}      
\centering          
\begin{tabular}{llccc}
\hline\hline                             
Source & Type      & RXTE Slew Survey &{\it  XMM-Newton}   & {\it XMM-Newton} Exposure Time \\
       &           & (cts/sec)$_{3-8~keV}$& (cts/sec)$_{2-10~keV}$ & (ksec)               \\
\hline                    
NGC4593       & BLSY &1.05 $\pm$ 0.16 &  4.613 $\pm$ 0.009 & 53.2\\
MRK704        & BLSY &1.06 $\pm$ 0.22 &  0.984 $\pm$ 0.009 & 14.9\\ 
ESO511-G030   & NCSY &1.10 $\pm$ 0.09 &  2.284 $\pm$ 0.005 & 76.2\\ 
NGC7213       & BLSY &1.12 $\pm$ 0.16 &  2.448 $\pm$ 0.009 & 29.6\\ 
AKN564        & NLSY &1.13 $\pm$ 0.02 &  2.246 $\pm$ 0.005 & 84.1\\
H1846-786     & NCSY &1.15 $\pm$ 0.09 &  0.700 $\pm$ 0.011 &  6.3\\ 
MRK110        & NLSY &1.17 $\pm$ 0.10 &  3.362 $\pm$ 0.010 & 32.8\\ 
ESO198-G024   & BLSY &1.21 $\pm$ 0.19 &  1.150 $\pm$ 0.003 & 85.3\\
FAIRALL9      & BLQ  &1.25 $\pm$ 0.02 &  1.313 $\pm$ 0.007 & 25.9\\
UGC3973       & BLSY &1.32 $\pm$ 0.12 &  1.304 $\pm$ 0.019 &  3.6\\
NGC4051       & NLSY &1.49 $\pm$ 0.02 &  2.783 $\pm$ 0.007 & 73.3\\ 
NGC526A       & NCSY &1.61 $\pm$ 0.14 &  2.177 $\pm$ 0.007 & 41.5\\
MCG-2-58-22   & BLSY &1.61 $\pm$ 0.08 &  3.459 $\pm$ 0.021 &  7.2\\
NGC7469       & BLSY &1.69 $\pm$ 0.01 &  3.343 $\pm$ 0.005 &142.6\\
MRK766        & NLSY &1.77 $\pm$ 0.12 &  1.962 $\pm$ 0.002 &398.3\\
MRK590        & BLSY &1.95 $\pm$ 0.10 &  0.766 $\pm$ 0.003 & 72.0\\ 
IRAS05078+1626& NCSY &2.08 $\pm$ 0.19 &  2.691 $\pm$ 0.008 & 40.1\\ 
NGC3227       & BLSY &2.10 $\pm$ 0.02 &  3.635 $\pm$ 0.006 & 97.4\\ 
MR2251-178    & BLQ  &2.10 $\pm$ 0.11 &  2.387 $\pm$ 0.007 & 44.5\\
MRK279        & BLSY &2.13 $\pm$ 0.06 &  3.110 $\pm$ 0.006 &104.8\\
ARK120        & BLSY &2.14 $\pm$ 0.03 &  4.553 $\pm$ 0.008 & 76.3\\ 
NGC7314       & NCSY &2.16 $\pm$ 0.04 &  4.509 $\pm$ 0.012 & 30.1\\
H0557-385     & NCSY &2.33 $\pm$ 0.34 &  4.254 $\pm$ 0.021 & 10.5\\
MCG+8-11-11   & BLSY &2.36 $\pm$ 0.27 &  4.838 $\pm$ 0.014 & 26.5\\
MCG-6-30-15   & BLSY &3.08 $\pm$ 0.02 &  4.719 $\pm$ 0.004 &295.1\\
MRK509        & BLQ  &3.12 $\pm$ 0.10 &  4.432 $\pm$ 0.005 & 27.8\\
NGC3516       & BLSY &3.21 $\pm$ 0.01 &  1.841 $\pm$ 0.004 &153.0\\
NGC5548       & BLSY &3.58 $\pm$ 0.02 &  4.191 $\pm$ 0.007 & 74.8\\
NGC3783       & BLSY &4.90 $\pm$ 0.04 &  5.539 $\pm$ 0.006 &170.4\\
MCG-5-23-16   & NCSY &6.46 $\pm$ 0.12 &  8.025 $\pm$ 0.008 &110.3\\
IC4329A       & BLSY &7.29 $\pm$ 0.07 & 10.713 $\pm$ 0.011 & 85.1\\ 
\hline
UGC10683      &   SY &1.27 $\pm$ 0.18 & $----$ & $----$ \\
ESO141-G055   &   SY &1.37 $\pm$ 0.12 & $----$ & $----$ \\
\hline                  
\end{tabular}
\tablefoot{
These sources, with the exception of the
last two sources, have all been the target of {\it XMM-Newton}
observations. The {\it XMM-Newton} EPIC-pn count rate in the 2-10~keV energy
range and exposure time is also given. For two of them, UGC~10683 and
ESO~0141-G055, no {\it XMM-Newton} data were available as of April 2008. This
reduces the number of XSS-selected bright sources in the FERO sample to 31.
}
\end{table*}

\section{List of sources with multiple observations that have not been summed.}\label{appendixb}
\par List of sources within the FERO sample where multiple observations are
available but only one has been used (the one with longest exposure time).
\begin{table}
\caption{List of sources within the FERO sample where multiple observations
are available but only one has been used (the one with longest exposure
time).}
\label{mult_observations_sing}
\centering          
\begin{tabular}{l c c c}     
\hline\hline                             
\multicolumn{1}{c}{Source} & List of     & Date & Exposure \\
       & {\it XMM-Newton}   &    & (ksec) \\ 
       & Observation &      &    \\ 
       & IDs combined&      &    \\ 
\hline 
1H0707-495              & {\bf 0148010301}&2002-10-13& 70.0\\ 
                        &      0110890201 &2000-10-21& 36.4\\ 
ESO113-G010             & {\bf 0301890101}&2005-11-10& 90.6\\ 
                        &      0103861601 &2001-05-03&  4.0\\ 
ESO198-G24              & {\bf 0305370101}&2006-02-04& 85.3\\ 
                        &      0067190101 &2001-01-24& 26.4\\ 
                        &      0112910101 &2000-12-01&  6.8\\ 
IZW1                    & {\bf 0300470101}&2005-07-18& 58.0\\
                        &      0110890301 &2002-06-22& 18.4\\
MR2251-178              & {\bf 0012940101}&2002-05-18& 44.5\\
                        &      0112910301 &2000-11-29&  3.5\\
MRK841                  & {\bf 0205340201}&2005-01-16& 30.1\\
                        &      0112910201 &2001-01-13&  5.9\\
                        &      0070740101 &2001-01-13&  7.6\\
                        &      0070740301 &2001-01-14&  9.3\\
                        &      0205340401 &2005-07-17& 17.3\\
NGC2622                 & {\bf 0302260201}&2005-04-09&  6.6\\ 
                        &      0302260401 &2005-10-08&  4.5\\
NGC4051                 & {\bf 0109141401}&2001-05-16& 73.3\\
                        &      0157560101 &2002-11-22& 46.2\\
NGC4593                 & {\bf 0059830101}&2002-06-23& 53.2\\
                        &      0109970101 &2000-07-02&  8.8\\
PG1211+143              & {\bf 0112610101}&2001-06-15& 49.5\\
                        &      0208020101 &2004-06-21& 43.6\\
UGC3973                 & {\bf 0400070201}&2006-09-30& 14.5\\ 
                        &      0103860801 &2000-10-09&  1.7\\
                        &      0103862101 &2001-04-26&  3.6\\
PG1116+215              & {\bf 0201940101}&2004-12-17& 69.3\\ 
                        &      0201940201 &2004-12-19&  5.0\\
                        &      0111290401 &2001-12-02&  5.5\\
IRASF12397+3333         & {\bf 0202180201}&2005-06-20& 68.7\\  
                        &      0202180301 &2005-06-23&  9.1\\ 
\hline                  
\end{tabular}
\tablefoot{
Observation IDs in bold correspond to those observations that have been
considered, while observation IDs not in bold have been discarded.
}
\end{table}

\section{Best-fit-model parameters for sources belonging to the flux-limited sample.}\label{appendixe}

\par Table \ref{tab_fls_model} gives some relevant best-fit-model
parameters (see Section~\ref{SpectralModel}) for the 31 sources belonging to
the flux-limited sample.

\begin{table*}
\caption{Flux-limited sample: relevant best-fit-model parameters corresponding
  to the neutral reflection run and 6.4~keV relativistic Fe K$_{\rm \alpha}$
  line.}\label{tab_fls_model}      
\centering
{\tiny
\begin{tabular}{lcccccccc}
\hline\hline                             
Source & Power Law &\multicolumn{2}{c}{Warm Absorber} & \multicolumn{2}{c}{Compton
  Reflection} & \multicolumn{3}{c}{EW  Narrow Fe K$_\alpha$ Emission
  Lines}\\
& $\Gamma$ & N$_{\rm H}$       & $\xi$ & R & Norm 
& Fe$_{I}$ 6.4~keV & Fe$_{XXV}$ 6.7~keV & Fe$_{XXVI}$ 6.96~keV \\
& & (10$^{22}$~cm$^{-2}$) & (erg~cm$^{-1}$~s$^{-1}$)  &   &  (10$^{-3}$~keV$^{-1}$~cm$^{-2}$~s$^{-1}$)
& (eV) & (eV) & (eV) \\
\hline                    
AKN564 &
2.82$_{-0.25}^{+0.05}$ &
0.76$_{-0.52}^{+0.17}$ &
$---$ &
2.22$_{-1.25}^{+0.62}$ &
18.38$_{-3.40}^{+1.36}$ &
$<$ 18&
13$^{+10}_{-11}$&
$<$ 4\\		    
NGC526A &
1.52$_{-0.02}^{+0.10}$ &
1.24$_{-0.05}^{+0.42}$ &
 $<$0.47 &
 $<$0.77 &
4.46$_{-0.21}^{+0.49}$ &
42$^{+13}_{-17}$&
$<$ 26	&
$<$ 22	\\	    
{\bf MCG-6-30-15} &
2.05$_{-0.01}^{+0.02}$ &
0.80$_{-0.14}^{+0.04}$ &
6.30$_{-2.98}^{+1.97}$ &
0.51$_{-0.17}^{+0.29}$ &
17.38$_{-0.49}^{+0.23}$ &
44$^{+4}_{-4}$&	
$<$ 5&
$<$ 7	\\	    
{\bf MRK766} &
2.10$_{-0.05}^{+0.06}$ &
0.40$_{-0.12}^{+0.19}$ &
10.37$_{-4.92}^{+8.71}$ &
0.78$_{-0.42}^{+0.77}$ &
 $>$6.87 &
32$^{+4}_{-6}$&
$<$ 21&
20$^{+9}_{-11}$\\	    
NGC5548 &
1.64$_{-0.01}^{+0.02}$ &
 $<$0.42 &
$---$ &
0.20$_{-0.12}^{+2.82}$ &
8.57$_{-0.04}^{+0.46}$ &
67$^{+7}_{-12}$	&
$<$ 4&
$<$ 7	\\	    
MCG+8-11-11 &
1.78$_{-0.04}^{+0.05}$ &
 $<$6.24 &
$---$&
1.37$_{-0.48}^{+0.54}$ &
11.78$_{-0.56}^{+0.62}$ &
97$^{+10}_{-11}$&
$<$ 10&
$<$ 17\\		    
MRK704 &
1.72$_{-0.14}^{+0.04}$ &
9.67$_{-3.86}^{+3.82}$ &
105.42$_{-30.54}^{+74.83}$ &
 $<$3.33 &
3.37$_{-0.84}^{+0.63}$ &
$<$ 83&
$<$ 42&
$<$ 45	\\	    
{\bf NGC3516} &
1.61$_{-0.06}^{+0.06}$ &
4.48$_{-0.52}^{+0.63}$ &
77.03$_{-16.48}^{+11.98}$ &
1.99$_{-0.32}^{+0.22}$ &
4.02$_{-0.27}^{+0.14}$ &
140$^{+9}_{-13}$&
$<$ 7&	
$<$ 5	\\	    
{\bf MCG-5-23-16} &
1.65$_{-0.02}^{+0.04}$ &
1.20$_{-0.02}^{+0.13}$ &
 $<$0.02 &
1.15$_{-0.18}^{+0.19}$ &
19.37$_{-0.48}^{+0.79}$&
32$^{+6}_{-6}$&	
$<$ 5 &
$<$ 5    \\
{\bf NGC3227} &
1.52$_{-0.02}^{+0.01}$ &
0.09$_{-0.03}^{+0.07}$ &
 $<$25.38 &
0.15$_{-0.12}^{+0.12}$ &
6.70$_{-0.19}^{+0.09}$ &
59$^{+9}_{-18}$	&
$<$ 5&
8$^{+6}_{-5}$\\	    
H0557-385 &
1.86$_{-0.05}^{+0.2}$ &
1.22$_{-0.32}^{+0.57}$ &
5.67$_{-5.24}^{+14.75}$ &
 $<$3.28 &
13.63$_{-2.52}^{+4.12}$ &
$<$ 31&			
$<$ 14&
$<$ 26\\		    
NGC4593 &
1.84$_{-0.04}^{+0.03}$ &
 $<$4.07 &
$---$&
1.24$_{-0.31}^{+0.32}$ &
11.88$_{-0.55}^{+0.44}$ &
39$^{+26}_{-16}$ &
$<$ 10 &
14$^{+8}_{-7}$	 \\ 
NGC7469 &
 $>$1.97 &
 $<$0.53 &
$---$&
1.79$_{-0.19}^{+0.25}$ &
10.07$_{-0.07}^{+0.20}$ &
73$^{+5}_{-6}$&
$<$ 3&
$<$ 9\\		    
{\bf NGC3783} &
1.70$_{-0.01}^{+0.01}$ &
1.64$_{-0.06}^{+0.08}$ &
54.21$_{-4.70}^{+4.09}$ &
0.51$_{-0.43}^{+0.04}$ &
13.44$_{-0.17}^{+0.08}$ &
90$^{+3}_{-8}$ &
$<$ 1 &	
23$^{+5}_{-3}$	\\    
{\bf ESO511-G030} &
1.93$_{-0.03}^{+0.02}$ &
 $<$1.38 &
$---$&
2.10$_{-0.10}^{+0.62}$ &
6.31$_{-0.06}^{+0.26}$ &
11$_{-0}^{+0}$ &
$<$ 14 &
$<$ 12 \\		    
IRAS05078+1626 &
1.72$_{-0.06}^{+0.03}$ &
4.94$_{-4.05}^{+3.00}$ &
$---$&
1.70$_{-0.76}^{+0.28}$ &
5.96$_{-0.38}^{+0.12}$ &
38$^{+21}_{-23}$&
11$^{+10}_{-10}$&
$<$ 11          \\ 
MRK279 &
1.83$_{-0.03}^{+0.02}$ &
 $<$0.51 &
$---$&
1.16$_{-0.35}^{+0.10}$ &
7.79$_{-0.07}^{+0.07}$ &
55$^{+16}_{-14}$&
$<$ 3	&
7$^{+7}_{-6}$\\	    
{\bf MRK509} &
 $>$1.96 &
 $<$0.2 &
$---$&
2.49$_{-0.36}^{+0.07}$ &
12.88$_{-0.18}^{+0.10}$ &
34$^{+4}_{-4}$&
$<$ 4	&
$<$ 3 \\		    
NGC7314 &
2.09$_{-0.08}^{+0.11}$ &
1.35$_{-0.29}^{+0.32}$ &
4.86$_{-3.81}^{+5.88}$ &
1.16$_{-0.64}^{+1.00}$ &
18.22$_{-2.02}^{+2.45}$ &
$<$ 56&
14$^{+10}_{-10}$&
27$^{+11}_{-11}$\\	    
NGC7213 &
1.76$_{-0.02}^{+0.09}$ &
 $<$7.67 &
$---$&
$<$0.77 &
5.80$_{-0.30}^{+0.38}$&
82$^{+14}_{-15}$&
22$^{+13}_{-13}$&
17$^{+13}_{-15}$\\	    
{\bf IC4329A} &
1.79$_{-0.01}^{+0.02}$ &
0.33$_{-0.02}^{+0.04}$ &
 $<$0.47 &
0.93$_{-0.21}^{+0.07}$ &
26.85$_{-0.37}^{+0.18}$&
34$^{+6}_{-9}$&
$<$ 4&	
8$^{+4}_{-3}$\\	    
MR2251-178 &
1.32$_{-0.01}^{+0.03}$ &
 $<$0.61 &
$---$&
0.54$_{-0.48}^{+0.24}$ &
3.06$_{-0.04}^{+0.10}$&
23$^{+11}_{-8}$	&
$<$ 5&
$<$ 2\\		    
H1846-786 &
1.87$_{-0.10}^{+0.21}$ &
 $<$8.95 &
$---$&
 $>$0 &
1.98$_{-0.19}^{+0.45}$&
78$^{+60}_{-58}$&
$<$ 35		&	      
104$^{+71}_{-72}$\\   
FAIRALL9 &
1.90$_{-0.11}^{+0.05}$ &
 $<$2.67 &
$---$&
1.70$_{-1.05}^{+0.65}$ &
3.56$_{-0.28}^{+0.22}$ &
107$^{+27}_{-23}$&
$<$ 21&			      
$<$ 29\\		    
UGC3973 &
1.78$_{-0.16}^{+0.27}$ &
 $<$37.79 &
$---$&
 $>$0 &
3.06$_{-0.59}^{+1.64}$ &
156$^{+57}_{-68}$ &
$<$ 87 &			      
$<$ 38 \\		    
MCG-2-58-22 &
1.69$_{-0.04}^{+0.10}$ &
 $<$2.93 &
$---$ &
$<$3.32 &
7.65$_{-0.49}^{+0.97}$ &
$<$ 51 &	
$<$ 20 &
21$^{+27}_{-19}$ \\
{\bf NGC4051} &
2.02$_{-0.01}^{+0.07}$ &
2.21$_{-0.64}^{+3.39}$ &
$---$ &
$<$1.49 &
9.48$_{-0.23}^{+1.09}$ &
63$^{+12}_{-13}$&
17$^{+10}_{-11}$&
$<$ 17	\\       
MRK590 &
1.61$_{-0.02}^{+0.08}$ &
 $<$3.39 &
$---$&
0.59$_{-0.39}^{+0.92}$ &
1.46$_{-0.04}^{+0.08}$ &
113$^{+16}_{-18}$&
$<$ 31		&	      
41$^{+20}_{-15}$\\
{\bf ARK120} &
2.12$_{-0.03}^{+0.02}$ &
 $<$0.52 &
$---$&
2.14$_{-0.34}^{+0.35}$ &
15.95$_{-0.34}^{+0.31}$ &
65$^{+7}_{-7}$&
15$^{+6}_{-6}$&
9$^{+7}_{-7}$\\
MRK110 &
1.88$_{-0.06}^{+0.02}$ &
 $<$1.21 &
$---$&
1.47$_{-0.61}^{+0.25}$ &
8.58$_{-0.39}^{+0.12}$ &
33$^{+10}_{-10}$ &
$<$ 6 &
$<$ 3 \\	
ESO198-G24 &
1.79$_{-0.04}^{+0.07}$ &
 $<$2.57 &
$---$&
1.79$_{-0.51}^{+0.83}$ &
2.7$_{-0.09}^{+0.16}$  &
60$^{+12}_{-15}$ &
$<$ 8&	
$<$ 8\\   
\hline\\
\end{tabular}
\tablefoot{
Errors are given at the 90\% confidence level. Upper limits to the
relevant parameters are also given at the 90\% confidence level. Sources
with a significant broad Fe K$_{\rm \alpha}$ line detection are marked in
bold (as listed in Table~\ref{tab_detections}).
}
}
\end{table*}

\section{Relativistic Fe K$_{\rm \alpha}$ line EW upper limits for the sources in the flux-limited sample.}\label{appendixd}

\par 
List of 20 sources within the flux-limited sample with an upper limit to the
relativistic Fe K$_{\rm \alpha}$ line EW. 

\begin{table*}
\caption{Flux-limited sample: list of 20 sources with an upper limit to
  the relativistic Fe K$_{\rm \alpha}$ line EW.}\label{FL_UL} 
\centering
\begin{tabular}{c c c}     
\hline\hline                             
Source &   Cts$^{2-10~keV}$   &   EW Upper Limit \\
       &    (10$^5$ cts)      &   (eV)\\ 
\hline 
         MRK704& 0.146 $\pm$ 0.012& 362.1\\ 
       FAIRALL9& 0.341 $\pm$ 0.018& 320.4\\ 
         AKN564& 1.889 $\pm$ 0.043& 292.3\\
    MCG-2-58-22& 0.250 $\pm$ 0.016& 208.5\\
      H1846-786& 0.044 $\pm$ 0.007& 173.7\\  
    MCG+8-11-11& 1.284 $\pm$ 0.036& 158.4\\ 
        NGC526A& 0.903 $\pm$ 0.030& 122.1\\  
         MRK110& 1.104 $\pm$ 0.033& 115.6\\
        NGC7213& 0.724 $\pm$ 0.027& 107.4\\ 
        UGC3973& 0.389 $\pm$ 0.020& 100.2\\ 
      H0557-385& 0.446 $\pm$ 0.021&  99.3\\ 
     MR2251-178& 1.062 $\pm$ 0.032&  99.2\\
        NGC7314& 1.358 $\pm$ 0.037&  87.8\\ 
        NGC4593& 2.455 $\pm$ 0.049&  84.2\\ 
        NGC7469& 4.769 $\pm$ 0.069&  84.0\\ 
 IRAS05078+1626& 1.079 $\pm$ 0.033&  61.4\\
         MRK279& 3.259 $\pm$ 0.057&  38.2\\ 
        NGC5548& 3.133 $\pm$ 0.056&  37.6\\  
     ESO198-G24& 0.981 $\pm$ 0.031&  35.8\\
         MRK590& 0.552 $\pm$ 0.023&  33.4\\ 
\hline                  
\end{tabular}
\tablefoot{
The EW upper limit is provided at the 90\% confidence level.
}
\end{table*}  

\section{A final self-consistent test on the detections within the flux-limited sample }\label{appendixf}

\par As a final test, the hard X-ray spectra of all the sources in the
flux-limited sample for which a significant detection of the relativistic
line can be claimed has been described with the most self-consistent
reflection model envisaged, but keeping the model as simple as
possible. The baseline model comprises one layer of ionised absorption (the
{\small ZXIPCF} model), reflection off cold distant matter including the most
important emission lines and the associated self-consistent reflection
continuum (the {\small PEXMON} model), and reflection from the accretion disc
(the \cite{Ross2005} {\small REFLION} model, convolved with the {\small
KYRLINE} kernel). Two ionised emission lines with energies fixed at 6.7~keV
and 6.96~keV are also included as in the previous phenomenological models used
throughout the paper, although they are not statistically required in all
cases.

\par The {\small PEXMON} model is described in detail in Nandra et al. (2007)
and describes the reflection spectrum from a cold slab of gas including both
the reflection continuum and the most relevant emission lines, which are
computed as self-consistently as possible according to the work by George \&
Fabian (1991). The metal abundances are fixed to the solar value (except for
the case of MCG--6-30-15, see below) and the reflector inclination with
respect to the line of sight to 60~degrees, which is appropriate for
torus-like reflection in Seyfert~1 galaxies. The illuminating continuum is a
power law with the same photon index as the power-law component of our
spectral model. The {\small REFLION} model describes reflection off a ionised
slab and is used instead to describe the disc reflection component. The
illuminating photon index is the same as the primary continuum and the Fe
abundance is fixed to the solar value (except for MCG--6-30-15, see
below). The model is convolved with the kernel of the {\small KYRLINE} model
(\cite{Dovciak2004}), which allows including all relativistic effects and
measuring the relevant parameters.

\par
It is important to stress once again that the goal here is not to provide the
best possible fitting statistics but rather to compare the results of the
phenomenological model used in the paper (which does not account for emission
lines and associated reflection continua in a self-consistent manner) with a
more physical spectral model. In some cases, more spectral components are
included as explained in the subsequent section.

\par
The results are reported in Table~\ref{testfinal}, where the most important
spectral parameters associated with the reflection components are
considered. In Table~\ref{FL_ModelTest1b} the best-fitting relativistic
parameters for both the phenomenological and the more physically motivated
models are reported. Such a comparison implies that the more complex and
self-consistent model for the reflection components does not significantly
affect the results. Since the cases considered here correspond to the highest
signal-to-noise data within the whole sample, the test supports the analysis
carried out on the whole available sample within the context of the less
sophisticated and more phenomenological model.

\begin{table*}
  \caption{Summary of the results of the self-consistent reflection
    model.}\label{testfinal} \centering
\begin{tabular}{lccccccc}     
\hline\hline                             
Source & Distant reflection   &  Disc reflection & Disc ionisation & $\theta$ & a & $\beta$ &
$\chi^2$/dof\\
 &    &   & (erg~cm~s$^{-1}$) & ($^\circ$) &  &  & \\

\hline 
IC4329A & $0.5\pm 0.2$ & $0.3 \pm 0.2$ & $150^{+80}_{-65}$ & $34^{+11}_{-12}$ & $\geq 0$& $2.0^{+1.3}_{-0.9}$ & 197.2/163 \\
MCG-5-23-16 & $0.3\pm 0.1$ & $0.5 \pm 0.3$ & $\leq 70$ & $31^{+7}_{-12}$ & $\geq 0$& $1.8^{+1.5}_{-0.8}$ & 247.3/164 \\
ESO511-G030 & $0.7\pm 0.2$ & $0.8 \pm 0.4$ & $70\pm 50$ & $37^{+21}_{20}$ & $\geq 0$& $2.6^{+0.8}_{-1.7}$ & 185.0/162 \\
MCG-6-30-15 & $0.2\pm 0.1$ & $2.4\pm 0.8$ & $62^{+40}_{-50}$ & $39\pm5$ & $\geq 0.8$ & 3.4$^{+0.8}_{-0.5}$ & 192.1/164 \\ 
NGC4051 & $0.9\pm 0.3$ & $1.0\pm 0.5$ & $140^{+90}_{-80}$ &$18^{+9}_{-13}$ & $\geq 0.3$ & 2.5$^{+0.6}_{-0.5}$ & 172.0/162 \\ 
NGC3516 & $1.5\pm 0.3$ & $1.8\pm 0.5$ & $\leq 70$ & $34^{+7}_{-12}$ &$\geq 0.3$ & 2.7$\pm 0.4$ & 212.7/161 \\ 
NGC3783 & $0.8\pm 0.2$ & $0.8\pm 0.4$ & $\leq 90$ & $27^{+19}_{-20}$& $\geq 0.2$ & 3.1$^{+0.5}_{-0.4}$ & 221.4/163 \\ 
NGC3227 & $0.6\pm 0.2$ & $0.5\pm 0.3$ & $70^{+60}_{-60}$ & $26\pm9$ &$\geq 0$ & 2.3$^{+1.1}_{-0.7}$ & 226.5/162 \\
MRK509 & $0.3\pm 0.1$ & $1.0\pm 0.6$ & $80^{+40}_{-50}$ & $\geq 45$ &$\geq 0.4$ & 2.8$^{+0.8}_{-0.5}$ & 244.2/160 \\
MRK766 & $0.3\pm 0.1$ & $0.8\pm 0.4$ & $700^{+120}_{-350}$ & $28\pm8$& $\geq 0.3$ & 2.6$^{+0.4}_{-0.5}$ & 262.5/162 \\ 
ARK120 & $1.0\pm 0.3$ & $0.7\pm 0.4$ & $80^{+80}_{-30}$ & $45\pm 10$ &$\geq 0.1$ & 2.7$^{+0.9}_{-1.2}$ & 188.7/160 \\ 
\hline                  
\end{tabular}
\tablefoot{
All sources for which a detection of a relativistic line
component is claimed in the paper have been described with two
self-consistent reflection models, one for distant reflection (the {\small
PEXMON} model), the other for disc reflection (the {\small REFLION}
model). The relevant parameters and fitting statistics are reported in
this table.
}
\end{table*}  

\begin{table*}
\caption{Comparison of the phenomenological and self-consistent model.}\label{FL_ModelTest1b} 
\centering
\begin{tabular}{l|ccc|ccc}     
\hline\hline                             
       & \multicolumn{3}{c|}{Phenomenological Model}   &
       \multicolumn{3}{c}{Self-consistent Model} \\
\hline
Source &  $\theta$  & a & $\beta$  &  $\theta$  & a & $\beta$ \\
       & ($^\circ$) &            &                & ($^\circ$)    &         &\\ 
\hline 
IC4329A     & 28$^{+6}_{-11}$ & $\geq 0$ & $\leq 1.3$ &34$^{+11}_{-12}$ & $\geq 0$ &  $1.9^{+0.8}_{-1.0}$ \\ 
MCG-5-23-16 & 21$^{+8}_{-3}$ & $\geq 0$ & $\leq 1.6$ & $31^{+7}_{-12}$ & $\geq 0$& $1.8^{+1.5}_{-0.8}$ \\ 
ESO511-G030 & $18\pm 7$  & $\geq 0$ & $\leq 1.1$ & $37^{+21}_{-20}$ & $\geq 0$& $2.6^{+0.8}_{-1.7}$ \\ 
MCG-6-30-15  & 40$^{+1}_{-3}$ & 0.86$^{+0.01}_{-0.02}$ & 4.1$\pm 0.2$ & $39\pm5$ & $\geq 0.8$ & 3.4$^{+0.8}_{-0.5}$ \\ 
NGC4051     & $22\pm 6$ & $\geq 0.46$ & 2.9$^{+0.3}_{-0.4}$ & 18$^{+9}_{-13}$ & $\geq 0.3$	&  2.5$^{+0.6}_{-0.5}$ \\ 
NGC3516     & 27$^{+2}_{-3}$ & $\geq 0.48$ & 2.8$^{+0.2}_{-0.3}$ & $34^{+7}_{-12}$ &$\geq 0.3$ & 2.7$\pm 0.4$ \\
NGC3783     & $\leq 8$ & $\geq 0.16$ & 2.7$^{+0.1}_{-0.2}$ &27$^{+19}_{-20}$ & $\geq 0.2$ &  3.1$^{+0.5}_{-0.4}$ \\
NGC3227     & $23\pm 4$ & $\geq 0$ & 1.9$^{+0.6}_{-0.5}$ &$26\pm 9$ & $\geq 0$ &  2.3$^{+1.1}_{-0.7}$ \\
MRK509   & $53 \pm 1$ & $0.78^{+0.03}_{-0.04}$ & $\geq 3.8$ & $\geq 45$ &$\geq 0.4$ & 2.8$^{+0.8}_{-0.5}$  \\
MRK766   & 20$^{+3}_{-2}$ & $\geq 0.47$ & 2.7$^{+0.2}_{-0.1}$ & $28\pm8$& $\geq 0.3$ & 2.6$^{+0.4}_{-0.5}$  \\ 
ARK120   & $\geq 59$ & $\geq 0 $ & 2.2$^{+0.6}_{-0.3}$ &$48\pm 12$& $\geq 0.1$ & 2.7$^{+0.9}_{-1.2}$ \\ 
\hline                  
\end{tabular}
\tablefoot{
This table shows a comparison of the best-fitting relativistic parameters (only
disc parameters are shown) obtained by
applying the phenomenological model discussed throughout the paper (whose
full results are given in Table~\ref{tab_detections}) and the more self-consistent one discussed in
Appendix~\ref{appendixf}. Here, only sources for which a detection of a relativistic line
component is claimed in the paper are shown.
}
\end{table*}  

\subsection{Notes on individual sources}

\par Slight modifications to the baseline model described in
Appendix~\ref{appendixf} are given here for some individual sources. In the
remaining cases, the baseline model was applied without being modified.\\
\\
{\bf IC4329A:} Absorption is best modelled with a neutral layer
with moderate column density ($\simeq$ 4 $\times$ 10$^{21}$~cm$^{-2}$).
A Gaussian absorption line with EW $\simeq$ -15~eV was also included to model
an absorption feature at $\simeq$ 7.65~keV phenomenologically.\\
\\
{\bf MCG-5-23-16:} Absorption is best modelled with a neutral layer
with column density N$_{\rm H} \simeq$ 1.5 $\times$
10$^{22}$~cm$^{-2}$. The presence of a further ionised layer is
possible but not statistically required.\\
\\
{\bf MCG-6-30-15:} The Fe abundance of the two reflectors was left
free to vary in this case resulting in an overabundance of $\sim
\times$ 3 with respect to the solar value. The model also comprises a
Gaussian absorption line at $\simeq$ 6.7~keV with EW $\simeq$ -15~eV.\\
\\
{\bf NGC4051:} A Gaussian
absorption line with EW$\simeq$ -30~eV was also included to model
an absorption feature at $\simeq$ 7.04~keV phenomenologically.\\
\\
{\bf NGC3516:}  A Gaussian
absorption line with EW$\simeq$ -28~eV was also included to model
an absorption feature at $\simeq$ 6.7~keV phenomenologically.\\
\\
{\bf NGC3783:} A Gaussian
absorption line with EW$\simeq$ -20~eV was also included to model
an absorption feature at $\simeq$ 6.7~keV phenomenologically.\\
\\
{\bf Mrk509:} A Gaussian
absorption line with EW$\simeq$ -10~eV was also included to model
an absorption feature at $\simeq$ 7.3~keV phenomenologically.\\
  
\end{appendix}

\end{document}